\documentclass{aa}

\usepackage{graphicx}
\usepackage{txfonts}
\usepackage{hyperref}
\usepackage{multicol}

\hypersetup{
    colorlinks=true,
    linkcolor=blue,
    filecolor=magenta,      
    urlcolor=cyan,
    citecolor=blue
}

\usepackage{comment}
\usepackage{placeins}
\begin{document} 

   \title{The efficiency of mixed modes for angular momentum transport}

   \author{B. Bordadágua \inst{1,}\inst{2},
          F. Ahlborn \inst{1},
          Q. Coppée \inst{1,}\inst{2},
           J. P. Marques \inst{3},
           K. Belkacem \inst{4},
          S. Hekker \inst{1,}\inst{2}
          }

   \institute{Heidelberger Institut fur Theoretische Studien (HITS), Schloss-Wolfsbrunnenweg 35, 69118 Heidelberg, Germany \\
    \email{beatriz.bordadagua@gmail.com}
    \and Center for Astronomy (ZAH/LSW), Heidelberg University, K\"onigstuhl 12, 69117 Heidelberg, Germany
    \and Université Paris-Saclay, CNRS, Institut d'astrophysique spatiale, 91405, Orsay, France
    \and LIRA, Observatoire de Paris, Université PSL, CNRS, Sorbonne Université, Université Paris Cité, 5 place Jules Janssen, 92195 Meudon, France
             }

\authorrunning{Bordadágua et al.}
   \date{Received month day, year; accepted month day, year}

  \abstract
   {Core rotation rates of red-giant stars inferred from asteroseismic observations are substantially lower than predicted by current stellar models. This indicates the lack of an efficient angular momentum transport mechanism in radiative interiors. Mixed pressure-gravity modes are a promising candidate to extract angular momentum from the core of red giants.}
   {We focus on determining the effect of mixed modes on the rotation rates of stars evolving along the red giant branch (RGB).} 
   {We developed a post-processing code that computes the angular momentum transport by meridional currents, shear-induced turbulence and mixed modes. Rotation rates were computed for models along the RGB with different stellar masses and different initial rotation profiles. }
   {We find that the mixed modes can explain some of the spin down observed in red giant stars, however, the values of non-radial mode amplitudes strongly affect the efficiency of this mechanism. Rotation rates from models neglecting radiative damping on the mixed mode amplitudes overlap with observations and produce a localised spin down around the hydrogen-burning shell, whereas the inclusion of radiative damping strongly suppresses and delays this spin down.
   We also show that including an additional viscosity term with values in the range $10^3-10^4\;\mathrm{cm}^2 \mathrm{s}^{-1}$ redistributes the localised spin down due to the mixed modes, enhancing their efficiency.}
   {Our results reveal that the mixed mode amplitudes need to be constraint to precisely quantify the spin down of red giant cores. Nevertheless, the mixed mode mechanism by itself cannot explain the full spread in observed core rotation rates along the RGB. This will only be possible with an additional mechanism for angular momentum transport. }

   \keywords{asteroseismology --
                stars: evolution --
                stars: interiors --
                stars: oscillations --
                stars: rotation
               }
   \maketitle

\section{Introduction}

\par Rotation plays a significant role in stellar evolution. It perturbs the equilibrium configuration of the star and modifies the stellar evolution. For example, rotation indirectly influences age estimates and the fate of a star \citep[e.g.,][]{2009pfer.book.....M}. Rotation also triggers a variety of processes such as mass loss linked to stellar winds, internal flow circulation and hydrodynamic instabilities. In fact, the chemical enrichment observed on the surface of massive stars may potentially be explained by instabilities triggered by rotational induced mixing \citep[e.g.,][]{2013EAS....62..227P}. In low mass stars, the differential rotation observed arises from a large redistribution of angular momentum between radiative and convective layers. The physical processes responsible for the transport of angular momentum in these stars remain an open question until today.

\par In the past decades, the CoRoT \citep{2006ESASP1306...33B} and \textit{Kepler} \citep{2010AAS...21510101B} space missions have provided a vast amount of data of solar-like oscillators \citep[see reviews by][]{2013ARA&A..51..353C,2013AdSpR..52.1581H,2016cole.book..197M,2017A&ARv..25....1H,2019LRSP...16....4G}. In red-giant stars the detected oscillations are stochastically excited by turbulent convection in the stellar envelope and can propagate to the innermost layers. The resulting standing waves, so-called mixed modes, behave as gravity modes in the radiative core and as acoustic modes in the envelope of the star. Thus providing important constraints on the structure of stellar interiors \citep[e.g.,][]{2011Sci...332..205B,2011Natur.471..608B,2011A&A...532A..86M}. In the presence of rotation, the degeneracy in the azimuthal order of non-radial modes is lifted. Consequently, frequency splittings associated with rotation can be observed in the non-radial modes in the power spectrum. The measurement of these rotational splittings in red giants allows us to infer the mean core and envelope rotation rates \citep[e.g.,][]{2012Natur.481...55B}. These measurements show that the core rotates around 10 times faster than the envelope. They have also revealed that core rotation remains roughly constant along the red giant branch (RGB) and seems independent of the stellar mass \citep[e.g.,][]{2012A&A...548A..10M,2018A&A...616A..24G}. 

\par In contrast with the observations, rotating stellar models in which one accounts for local conservation of angular momentum predict that the core should rotate 100 to 1000 times faster than the envelope \citep[e.g.,][]{2014ApJ...788...93C}.
Whilst these models show that the rotation evolution along the RGB is dominated by the contracting core, observations of core rotation rates \citep{2012Natur.481...55B,2012A&A...548A..10M,2014A&A...564A..27D,2016ApJ...817...65D,2017A&A...602A..62T,2018A&A...616A..24G,2020A&A...641A.117D,2024A&A...688A.184L} point towards a physical mechanism that is counteracting the spin up due to contraction and slowing down the stellar core. Therefore, an efficient mechanism to transport angular momentum from the radiative core to the outer layers is needed. 

\par Stellar models that include torques solely generated by hydrodynamic processes, such as meridional circulation and shear instabilities \citep{1992A&A...265..115Z,1998A&A...334.1000M}, are also inefficient at slowing down the cores of red giants \citep{2000ApJ...528..368H,2012A&A...544L...4E,2013A&A...549A..74M,2013A&A...555A..54C} and fail to reproduce the small degree of differential rotation found in the main-sequence progenitors of red-giant stars \citep{2019A&A...626A.121O,2023A&A...677A...6M}. Other physical processes have to be considered, and several mechanisms have been proposed to explain the efficient redistribution of angular momentum in stellar interiors \citep[see extensive reviews by][]{2013EAS....62..227P,2019ARA&A..57...35A,2024arXiv240911354E}.

\par In particular, mechanisms including internal magnetic fields have been successful in reproducing red giant rotation rates. One example is large-scale fossil magnetic fields \citep{2015ApJ...808...35K,2021A&A...646A..19T}. They enforce rigid rotation in radiative zone and differential rotation in the envelope, thus contradicting some recent studies \citep{2017MNRAS.464L..16K,2018ApJ...862....9D}. Another possibility is the recent revision of the magnetohydrodynamic instability known as the Tayler-Spruit dynamo \citep{2002ApJ...574L.175T,2019MNRAS.485.3661F,2022A&A...664L..16E} or the azimuthal magneto-rotational instability \citep{2016A&A...589A..23S,2023A&A...673A.110M}.
These magnetic instabilities reproduce the nearly uniform rotation profile observed in the main-sequence progenitors of red-giant stars \citep[i.e., gamma Doradus stars;][]{2020A&A...640A..49O,2021MNRAS.502.5856S,2023A&A...673A.110M,2024A&A...681L..16M} and a differentially rotating core consistent with observations in the RGB \citep{2019MNRAS.485.3661F}. However, it is important to note that the efficiency of these mechanisms for angular momentum transport depends on the calibration of free parameters for the different evolutionary stages \citep{2019A&A...631L...6E,2020A&A...634L..16D,2022A&A...664L..16E,2023A&A...673A.110M,2024A&A...681L..16M}. This may potentially hide the need for additional physical processes.

\par Another physical mechanism to consider is transport through waves. Internal gravity waves (IGW) are buoyancy (gravity) waves that propagate in stably-stratified radiative regions. They can originate from turbulent stresses or from penetrative convection at the base of the convection zone. IGWs can generate a flux of energy, angular momentum, and chemical mixing within the radiation zone. Despite explaining the nearly rigid rotation profile in the radiative zone of the Sun \citep{1999ApJ...520..859K,2002ApJ...574L.175T,pincon_generation_2016}, both \cite{2014ApJ...796...17F} and \cite{pincon_implications_2017} have shown that strong radiative damping in the hydrogen burning shell prevents these waves from travelling to the innermost layers in more evolved stars. Thus inhibiting their ability to spin down the cores of stars ascending the RGB, mainly contributing to the angular momentum redistribution in the subgiant phase.

\par Mixed modes on the other hand consist of standing waves that propagate through convective and radiative regions enabling energy exchanges with the mean flow and consequently extracting angular momentum from the radiative zone \citep{2015A&A...579A..31B,2015A&A...579A..30B}. The work of \cite{2015A&A...579A..31B} discussed the efficiency of mixed modes for transport of angular momentum at different stages of evolution. For the subgiant and early red giant phase the angular momentum extracted by mixed modes is not enough to counterbalance the spin up of the stellar core due to contraction. Later in the red giant phase, due to the increase of the number of mixed modes, their amplitude and wave number, the mixed modes are able to spin down the hydrogen burning shell. The timescales associated with the transport of angular momentum by mixed modes suggest that the efficiency of this mechanism increases along the RGB, highlighting its potential to efficiently spin down the core of more evolved red-giant stars \citep{2015A&A...579A..31B}.

\par Despite the evidence from the different timescales, some open questions concerning the angular momentum transport by mixed modes remain. For example, whether the mixed mode mechanism is sufficient to slow down the entire core of red giants. Another unknown is how the increasing radiative damping as the star evolves along the RGB affects the mode amplitudes and the efficiency of this mechanism. Moreover, until now the hypothesis of angular momentum transport by mixed modes has not been tested in stellar evolution codes. 
It is likely that different angular momentum transport mechanisms are at work at different stages of evolution or simultaneously. Hence, future efforts are needed to test all these theories and 
take into account the possible interactions between these mechanisms.

\par Several stellar evolutionary codes already include angular momentum transport by hydrodynamic instabilities and magnetic fields through a diffusive approach -- for example in MESA \citep[][]{paxton_modules_2013} -- or by an advective-diffusive approach -- as in STAREVOL \citep{palacios_rotational_2003}, Geneva \citep{1997A&A...321..465M,2008Ap&SS.316...43E} and CESTAM \citep{2013A&A...549A..74M}. The inclusion of diffusive and advective angular momentum transport processes along stellar evolution poses a real challenge to computational models due to the wide variety of spatial and temporal scales \citep[e.g.,][]{2013EAS....62..227P}. For the purpose of determining the effect of mixed modes on the rotation profile of red giant stars, we followed the advective-diffusion approach described in \cite{2013A&A...549A..74M}.

\par In this work, we computed rotation profiles and their evolution along the RGB including angular momentum transport by mixed modes. For this purpose, we developed a post-processing code that solves the angular momentum transport equation including turbulent diffusion coefficients and advection by meridional circulation and the mixed modes momentum flux along evolution. We demonstrate that mixed modes can explain part of the spin down observed in red giant stars along evolution. However, the choice for non-radial mode amplitudes and their damping rates strongly affects the efficiency of this mechanism. We also show that another mechanism for angular momentum transport has to be taken into account to redistribute the localised spin down due to the mixed modes and explain the full spread in observed core rotation rates.

\par The paper is organized as follows: in Section~\ref{sec:2.1}, we recall the mixed modes theoretical formalism and discuss two approaches to estimate mode amplitudes. In Section~\ref{sec:2}, we describe the physical inputs included in our models and the numerical implementation. In Section~\ref{sec:timescales}, we describe the characteristic timescales for angular momentum transport by different mechanisms and discuss their efficiencies. Finally, in Section~\ref{sec:Quantitative estimate of spin down of red giants cores by mixed modes}, we show the evolution of rotation rates against asteroseismic observations. In Section~\ref{sec:Testing the efficiency of the mixed mode mechanism with different physics}, we present tests of different initial rotation profiles and include an additional diffusion mechanism. Section~\ref{sec:6} is dedicated to the conclusions and final remarks.

\section{Angular momentum transport}
\label{sec:2.1}

In this section we recall the formalism established by \cite{2015A&A...579A..30B} that describes how mixed modes transport angular momentum in the radiative region of low-mass evolved stars (Sect.~\ref{sec:wavemomentumflux}). Since the flux of angular momentum due to mixed modes highly depends on the non-radial mode amplitudes, we describe how we estimated radial and non-radial mode amplitudes (Sect.~\ref{sec:modeamplitudes}).

\subsection{Angular momentum flux due to waves}
\label{sec:wavemomentumflux}

\par Here we briefly describe the steps required to explain the theoretical formalism behind the angular momentum transported by mixed modes and refer the reader to the work of \cite{2015A&A...579A..31B,2015A&A...579A..30B} for further details.

\par \cite{2015A&A...579A..30B} demonstrated that mixed modes can transport angular momentum due to the exchange of momentum between the oscillations and the mean flow. The fluid in the interior of a star can be decomposed into a mean flow and perturbations to the mean flow, where the perturbations correspond to non-radial oscillations. Introducing this decomposition in the continuity, momentum, and energy conservation equations allows us to compute the effect of waves on the mean flow.
Subsequently, one needs to isolate the terms reflecting the effect of waves on the energy equation (hereafter called the wave heat flux), and on the momentum equation (hereafter called the wave momentum flux).
To gather both these fluxes into a single term one makes use of the transformed Eulerian mean formalism \citep[introduced by][]{1976JAtS...33.2031A,1978JAtS...35..175A}, showcasing how both fluxes are coupled to the meridional circulation and cannot be neglected. 

\par The novelty of the \citet{2015A&A...579A..30B} formalism is the inclusion of the wave heat flux into the momentum equation. Previous works have neglected this, which leads to inaccurate and imprecise estimates of the effect of the waves on the mean flow and consequently to inaccurate and imprecise estimates for the capability of waves to transport angular momentum.

\par The approximation of shellular rotation is valid under the assumption that in radiative zones the turbulence in the horizontal direction is much higher than in the vertical direction \citep{1992A&A...265..115Z}. With this approximation the angular velocity $\Omega$ of a rotating star is assumed constant on an isobar. Hence at dominant order, the mean angular momentum transport equation results in
\begin{equation}
\label{eq:AM}
   \langle \rho\rangle \frac{\mathrm{d} (r^2 \Omega)}{\mathrm{d}t}  = -\frac{1}{r^2}\frac{\partial}{\partial r} \left[ r^2 \left(\mathcal{F}_{\mathrm{shear}}+\mathcal{F}_{\mathrm{circ}}+\mathcal{F}_{\mathrm{waves}}  \right)\right],
\end{equation}
where $\rho$ is the density, $r$ is the stellar radius, the brackets $\langle .\rangle$ denote the horizontal (i.e. meridional and azimuthal) average, and $\mathrm{d}/\mathrm{d}t = \partial/\partial t + \dot{r} \partial/\partial r$ is the Lagrangian derivative. The fluxes write as follows
\begin{align}
\label{eq:AMshear}
    &\mathcal{F}_{\mathrm{shear}} =  - \langle \rho\rangle r^2 \nu_{\mathrm{v}} \frac{\partial \Omega}{\partial r}\;, \\
\label{eq:AMcirc}
    &\mathcal{F}_{\mathrm{circ}} = - \frac{1}{5}\langle \rho\rangle r^2 \Omega U_2 \;, \\
\label{eq:AMwaves}
    &\mathcal{F}_{\mathrm{waves}} =  \langle \rho\rangle \left\langle r \sin{\theta} \left[ \overline{\varv_{\phi}' \varv_r'} + 2 \cos{\theta}\; \Omega\; \overline{\varv_{\theta}' s'}\; \left(\frac{\mathrm{d} \langle s\rangle}{\mathrm{d} r} \right)^{-1} \right] \right\rangle   \;. 
\end{align}
The shear turbulent stress is computed using the vertical (i.e. radial) component of an eddy viscosity $\nu_{\mathrm{v}}$ \citep{1997A&A...317..749T,2004A&A...425..243M}. The vertical component of the meridional circulation velocity $U_2$ follows \cite{1998A&A...334.1000M}. The $\varv_r$, $\varv_{\phi}$ and $\varv_{\theta}$ are the radial, azimuthal, and meridional component of the wave velocity field, respectively, and $s$ is the specific entropy. The overbar denotes the azimuthal average and prime marks the perturbations with respect to the azimuthal average corresponding to non-radial oscillations. 

\par The explicit expression for the wave flux (Eq.~\ref{eq:AMwaves}) arises from the perturbation of the azimuthal and radial flow due to the propagation of waves (first term) and from incorporating the advective part of the wave heat flux term into the meridional flow (second term).

\par The wave velocity field in the radiative layers of low-mass evolved stars was modelled assuming the asymptotic, quasi adiabatic and slow-rotation ($\sigma_R>>\Omega$, where $\sigma_R$ is the real part of the frequency) approximations and computed through the non-radial non-adiabatic oscillation equations in the presence of rotation \citep{1989nos..book.....U}. After some rewriting one obtains the divergence of the wave flux, $\dot{\mathcal{J}}$, in the momentum Equation~\eqref{eq:AM}:
\begin{align}
\label{eq:j_coeffs}
     &\Dot{\mathcal{J}} \equiv -\frac{1}{r^2}\frac{\partial}{\partial r} \left[ r^2 \mathcal{F}_{\mathrm{waves}}\right] = \nonumber \\
     &\sum_{l,m} a^2_{l,m} \left(\mathcal{A}_{l,m}\frac{\partial^2\left(r^2 \Omega \right)}{\partial r^2}  + \mathcal{B}_{l,m}\frac{\partial\left(r^2 \Omega \right)}{\partial r} + \mathcal{C}_{l,m} \Omega + m \hat{\sigma}\mathcal{D}_{l,m}\right) \,,
\end{align}
with
\begin{align}
\label{eq:omega_R}
    \hat{\sigma} = \sigma_R + m\Omega\;,
\end{align}
where $a_{l,m}$ is the amplitude of a mode with angular degree $l$ and azimuthal order $m$, and, the coefficients $\mathcal{A}_{l,m}$, $\mathcal{B}_{l,m}$, $\mathcal{C}_{l,m}$ and $\mathcal{D}_{l,m}$ are given by Eqs (A.25 - A.28) of \cite{2015A&A...579A..30B}.

\par The first term of the coefficient $\mathcal{D}_{l,m}$ in Equation~\eqref{eq:j_coeffs} originates from the second term in Equation~\eqref{eq:AMwaves} and contains the dominant contribution to the angular momentum transport \citep{2015A&A...579A..30B}, exceeding the other terms by a few orders of magnitude. Therefore, the overall contribution of the mixed modes to the angular momentum transport arises from the wave heat flux. The expression for the divergence of the wave flux can be simplified to,
\begin{align}
\label{eq:j_d}
    &\Dot{\mathcal{J}} \approx \sum_{l,m} a^2_{l,m} m \hat{\sigma} \;\mathcal{D}_{l,m} \;\;\;\;\;\;\mathrm{with}\\
\label{eq:Dlm}
    &\mathcal{D}_{l,m} = \left[k_r^2  \left|\xi_r^{l,m}\right|^2 \rho\alpha \left(1-\frac{N^2}{\hat{\sigma}^2}\right) + \mathrm{non\;dominant\;terms}\right] \;,
\end{align}
where $\xi_r^{l,m}$ is the radial displacement of the wave, $N$ and $S_l$ are the buoyancy and Lamb frequencies, respectively. The radial wave number $k_r$ follows
\begin{align}    
    &k_r^2 = \frac{\sigma_R^2}{c_s^2} \left(1- \frac{S_l^2}{\sigma_R^2} \right) \left(1- \frac{N^2}{\sigma_R^2} \right) \;,
\end{align}
with $c_s$ the sound speed. The factor $\alpha$ is defined as
\begin{align}
    &\alpha = - \frac{L}{4\pi r^2 \rho T} \left( \frac{\nabla_{\mathrm{ad}}}{\nabla} - 1 \right) \left(\frac{d s}{dr}\right)^{-1}\;,
\end{align}
with $L$ and $T$ the luminosity and temperature, $\nabla$ and $\nabla_{\mathrm{ad}}$ the temperature and adiabatic gradient.

\par As discussed in \citet{2015A&A...579A..30B}, the prograde modes ($m<0$) extract angular momentum from the radiative region thus slowing down the core,
\begin{align}
\label{eq:prograde}
    \dot{J}_{l,-|m|} = -a^2_{l,-|m|} |m| \hat{\sigma} \;\mathcal{D}_{l,-|m|} \;,
\end{align}
whereas the retrograde modes ($m>0$) increase the angular momentum in the radiative region thus spinning up the core,
\begin{align}
\label{eq:retrograde}
    \dot{J}_{l,|m|} = a^2_{l,|m|} |m| \hat{\sigma} \;\mathcal{D}_{l,|m|} \;,
\end{align}
where $\dot{J}_{l,m}$ denotes the momentum flux for one mixed mode with angular degree $l$ and azimuthal order $m$. Without rotation, the net angular momentum flux is zero, since the prograde and retrograde modes contribution to the angular momentum cancels out. With rotation, the Doppler shift in the frequencies (Eq.~\ref{eq:omega_R}) causes a difference in the absolute value of the momentum flux of prograde and retrograde modes (Eq.~\ref{eq:prograde} and \ref{eq:retrograde}). Therefore, the resulting net momentum is given by,
\begin{align}
     \dot{J}_{l,-|m|} + \dot{J}_{l,|m|} \approx a^2_{l,|m|} \frac{2|m|^2}{\sigma_R^2} k_r^2  \left|\xi_r^{l,|m|}\right|^2  \;\rho \alpha N^2 \Omega \;,
     \label{eq:j_dot_}
\end{align}
where we inserted $\mathcal{D}_{l,m}$ from Equation~\eqref{eq:Dlm} into Equation~\eqref{eq:prograde} and Eq.~\eqref{eq:retrograde}, summed the contributions, expanded in $\Omega/\sigma_R$ and kept the leading term. This expression reveals that the combined effect of mixed modes is then to extract angular momentum from the radiative core (where Eq.~\ref{eq:j_dot_} yields negative values) and increase the angular momentum in the envelope (where Eq.~\ref{eq:j_dot_} yields positive values), thus slowing down the core of red giants. 

\par Accounting for all the mixed modes in a frequency range, the total mixed modes momentum flux yields,
\begin{align}
     \dot{\mathcal{J}} \approx \sum_{l} \sum_{|m|} a^2_{l,|m|} \frac{2|m|^2}{\sigma_R^2} k_r^2  \left|\xi_r^{l,|m|}\right|^2  \;\rho \alpha N^2 \Omega \,. 
     \label{eq:j_dot}
\end{align}

\par The efficiency of angular momentum extraction strongly depends on the mode amplitude $a_{l,m}$. The relation between the mode amplitude and the surface velocity $V_{l}$ is given by
\begin{align}
    a^2_{l,m} = \frac{2 V_l^2}{\sigma_R^2 \left|\mathbf{\xi}^{l,|m|} (R)\right|^2} \;,
    \label{eq:a_lm}
\end{align}
\citep{2011LNP...832..305S} where $\mathbf{\xi}^{l,|m|}(R)$ is the total wave displacement vector evaluated at the surface of the star. Hence, a quantitative estimate of the momentum flux (Eq.~\ref{eq:j_dot}) requires accurate estimates of surface velocities for non-radial modes (radial modes and modes with $m=0$ have a zero net momentum flux) which we will introduce in the next subsection.

\subsection{Mode surface velocities}
\label{sec:modeamplitudes}
\par We first establish radial mode surface velocities in Section~\ref{sec:radialmode}. Then we use them to compute non-radial mode surface velocities in Section~\ref{sec:nonradialmodeamplitudes} \citep[e.g., see][]{2015A&A...579A..31B}.

\subsubsection{Radial mode amplitudes}
\label{sec:radialmode}

\par
\begin{figure}
   \raggedright
   \includegraphics[width=9cm]{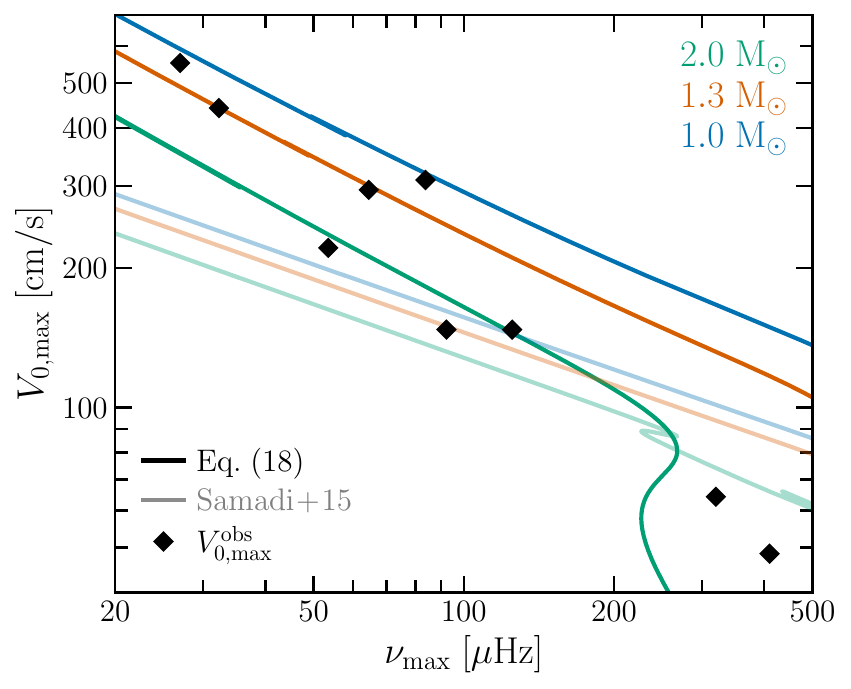}
   \caption{Radial mode surface velocities (logarithmic scale) as a function of the frequency at maximum oscillation power, $\nu_{\mathrm{max}}$ (logarithmic scale). The fully-opaque lines represent models evaluated using Eq.~\eqref{eq:V_max}. The semi-transparent lines show models evaluated using Eq.~(5.21) from \citet{2015EAS....73..111S}. The colour represent the masses of the evolutionary tracks. The black diamonds show the ground-based Doppler velocity measurements compiled by \cite{2012A&A...543A.120S}.} 
    \label{fig:Vmax_numax}
\end{figure}

 Previous works have derived scaling relations for the radial mode surface velocities $V_{0,\mathrm{max}}$ (at the frequency of maximum oscillation power $\nu_{\mathrm{max}}$) that match observations of main-sequence stars reasonably well \citep[for more details see, e.g.,][]{2011LNP...832..305S,2012A&A...543A.120S,2013EPJWC..4303008S,2015EAS....73..111S}. However, for red giants these scaling relations underestimate $V_{0,\mathrm{max}}$ compared with the few radial velocity measurements available (Doppler velocity measured from the ground -- see semi-transparent lines in Fig.~\ref{fig:Vmax_numax}). 

\par The amount of data available from \textit{Kepler} photometric observations allows us to better constrain radial mode amplitudes. Hence, in the present work we have adopted an empirical scaling relation obtained by \cite{2018A&A...616A..94V} for the radial mode bolometric amplitude $A_{0}^{\mathrm{bol}}$ of roughly 5000 stars,
\begin{align}
    \frac{A_{0}^{\mathrm{bol}}}{A_{0,\odot}^{\mathrm{bol}}} =  \left(\frac{L}{L_{\odot}} \right)^{0.61} \;\left(\frac{M}{M_{\odot}} \right)^{-1.14} \;\left(\frac{T_{\mathrm{eff}}}{T_{\mathrm{eff,}\odot}} \right)^{-5.48} \;,
    \label{eq:V_max_bol}
\end{align}
where $A_{0,\odot}^{\mathrm{bol}} = 2.53\pm0.11$ ppm corresponds to the maximum solar bolometric mode amplitude \citep{2009A&A...495..979M,2013EPJWC..4303008S}, $M$ is the stellar mass, and the solar reference values are indicated with the symbol $\odot$. 
Due to the importance of the effective temperature, $T_{\mathrm{eff}}$, compared with $L$ and $M$, we adopted the exponents of the fit including $T_{\mathrm{eff}}$ despite the large uncertainty in the exponent reported by \cite{2018A&A...616A..94V}. The values of these exponents are larger than the ones obtained by \cite{2012A&A...543A.120S}, bringing theoretical radial mode amplitudes closer to observations. We note that \cite{2018A&A...616A..94V} only reported the exponents of the fit, hence the solar reference values we adopted can lead to slightly different values for the radial mode bolometric amplitude than the ones reported in \cite{2018A&A...616A..94V}.

\par Subsequently, we used the non-adiabatic relation between bolometric mode amplitudes and mode surface velocities from \cite{2013EPJWC..4303008S}:
\begin{align}
    V_{0,\mathrm{max}} = V_{\odot,\mathrm{max}}\;\frac{A_{0}^{\mathrm{bol}}}{A_{0,\odot}^{\mathrm{bol}}} \; \left[\zeta_0 \; \left(\frac{L}{L_{\odot}} \right)^{0.25} \; \left(\frac{M}{M_{\odot}} \right)^{-0.25} \right]^{-1}\;,
    \label{eq:zeta_nonad}
\end{align}
where $V_{\odot,\mathrm{max}}=18.5\pm1.5$ cm/s is the maximum of the mode (intrinsic) surface velocity inferred at the photosphere \citep{2010A&A...509A..16S} and $\zeta_0 =0.59\pm0.07$ is the non-adiabatic coefficient estimated by \citet{2012A&A...543A.120S}. The latter was obtained using the MAD non-adiabatic pulsation code \citep[which includes time-dependent convection as described in][]{2005A&A...434.1055G} and brings the scaling relations closer to observations, contrary to an adiabatic coefficient.

\par Combining Equation~\eqref{eq:V_max_bol} with Equation~\eqref{eq:zeta_nonad} we obtained the following scaling relation to compute radial mode surface velocities,
\begin{align}
    V_{0,\mathrm{max}} = V_{\odot,\mathrm{max}} \; \zeta_0^{-1} \; \left(\frac{L}{L_{\odot}} \right)^{0.36}  \; \left(\frac{M}{M_{\odot}} \right)^{-0.89}\; \left(\frac{T_{\mathrm{eff}}}{T_{\mathrm{eff,}\odot}} \right)^{-5.48}\;,
    \label{eq:V_max}
\end{align}
which is illustrated with filled lines in Figure~\ref{fig:Vmax_numax}. The frequency ($\nu$) dependence of radial mode surface velocities $\left(V_{0,\mathrm{max}} \longrightarrow V_{0}(\nu)\right)$ is then taken into account following a Gaussian envelope \citep{2010A&A...517A..22M,2015A&A...579A..31B}.

\subsubsection{Non-radial mode amplitudes}
\label{sec:nonradialmodeamplitudes}

\par Non-radial mode surface velocities, $V_l$, can now be estimated from the radial mode surface velocities using the ratio,
\begin{align}
    \left( \frac{V_l(\nu)}{V_{0}(\nu)} \right)^2 \approx \left(\frac{\mathcal{M}_{0}}{\mathcal{M}_l} \right)^2 \frac{\Gamma_{0}}{\Gamma_l} \approx \left(\frac{\mathcal{M}_{0}}{\mathcal{M}_l} \right)^2  \frac{\eta_{0}}{\eta_l} \;,
    \label{eq:V_l}
\end{align}
(for details see Eq.~8 from \cite{2014ApJ...781L..29B}\footnote{The $V_l$ in our paper represents the mode surface velocity, which is a different notation than in \cite{2014ApJ...781L..29B}.} and Eq.~14 from \cite{2015A&A...579A..31B}) where  $\mathcal{M}_l$ is the mode mass, and $\Gamma_l$ the mode linewidth related to the mode damping rate $\eta_l$ by $\Gamma_l = \eta_l/\pi$.

\par The damping rate of a mode can be estimated by computing the work $\left(\int \mathrm{d}W\right)$ performed by the gas over one oscillation period and is given by 
\begin{align}
    \eta_l = -\frac{\int \mathrm{d}W}{2\pi \sigma_R \mathcal{M}_l \left|\xi_r^l(R)\right|^2} \;,
    \label{eq:eta}
\end{align}
\citep[see more details, e.g.,][]{1989nos..book.....U, 2009A&A...506...57D, 2018A&A...620A..43D} 
where $\mathcal{M}_l=\int_0^M \left|\mathbf{\xi}^l\right|^2 / \left|\xi_r^l(R)\right|^2\mathrm{d}m$ is the mode mass. 

\par In red giants, the work integral can be independently evaluated in the convective outer envelope and in the radiative core. Hence, the damping rate of a mixed mode can be written as \citep{2017A&ARv..25....1H}
\begin{align}
    \eta_l = \frac{\mathcal{M}_{\mathrm{env}}}{\mathcal{M}_l} \eta_{\mathrm{env}} + \frac{\mathcal{M}_{\mathrm{core}}}{\mathcal{M}_l} \eta_{\mathrm{core}}\;.
    \label{eq:eta_l}
\end{align}

\par The work performed within the convective region comes from stochastic excitation by turbulent motions and from damping by several different processes such as turbulent pressure, convective flux variations, and dissipation of kinetic energy \citep{1999A&A...351..582H}. To compute mode damping rates in the non-adiabatic outer layers requires non-adiabatic computations of oscillations and their interaction with convection \citep[see more details in][]{2015LRSP...12....8H}. Time-dependent treatment of convection has been introduced when solving the non-adiabatic pulsation equations by \cite{2009A&A...506...57D} and \cite{2014A&A...572A..11G} still with some limitations. 

\par To minimize the computation time for each model, in this work we abstain from evaluating the work integral in the outer layers of the star and instead resort to using scaling relations obtained from observations as was also done by \citet{2015A&A...579A..31B}. Therefore, to estimate the damping rate in the outer layers we use the following expression,
\begin{align}
\label{eq:belkacem+12}
    \eta_{\mathrm{env}} = \eta_{\odot} \; \left(\frac{T_{\mathrm{eff}}}{T_{\mathrm{eff},\odot}}\right)^{10.8} \left(\frac{\mathnormal{g}_{\mathrm{surf}}}{g_{\mathrm{surf},\odot}}\right)^{-0.3} \;,
\end{align}
\citep{2012A&A...540L...7B} where $\eta_{\odot} = 2.98\;\mathrm{rad/s}$ and $g_{\mathrm{surf}}$ is the surface gravity of the star.  
This expression shows the strong dependence of the damping rates on the effective temperature, however it neglects the frequency dependence we see in observations. 

\par In radiative regions, the work results from the heat provided to the gas by radiation during one oscillation period \citep{2009A&A...506...57D}. Without the contribution of the turbulent pressure and assuming the eigenfunctions to be real \citep[see more details in][]{2017A&ARv..25....1H}, the work in the core in Equation~\eqref{eq:eta} is reasonably well modelled by the asymptotic formulation of \cite{2001MNRAS.328..601D} and therefore the damping rate in the core is given by,
\begin{align}
    \eta_{\mathrm{core}} = \frac{\left[l(l+1) \right]^{3/2}}{8\pi \sigma_R^3 \int_0^{R_c} k_r \mathrm{d}r} \int_0^{R_c} \frac{\nabla_{\mathrm{ad}}-\nabla}{\nabla} \frac{\nabla_{\mathrm{ad}}N g L}{P r^5} \mathrm{d}r\;,
    \label{eq:eta_core}
\end{align}
where $g$ is the local gravity and $R_c$ is the radius at the base of the convection region. 
This expression reveals that the core damping is smaller in earlier phases and increases as the star evolves along the RGB as the luminosity, gravity and buoyancy frequency increase. \cite{2017A&ARv..25....1H} have shown that radiative damping in the core around the RGB bump would likely affect the observed properties of the modes since the mode damping time becomes of the same order of the duration of the nominal \textit{Kepler} mission. This expression also reveals that the radiative damping has a stronger contribution around the hydrogen burning shell where there is a peak in the buoyancy frequency. This is the region where the mixed modes are most efficient at transporting angular momentum as reported by \cite{2015A&A...579A..31B,2015A&A...579A..30B}. Hence, a precise estimate of the mode damping rate is required to properly infer the efficiency of the mixed mode mechanism for angular momentum transport. 

\par Finally, by introducing Equation~\eqref{eq:eta_l} into Equation~\eqref{eq:V_l}, the non-radial mode surface velocity, $V_{l}$, is given by
\begin{align}
    \left( \frac{V_l(\nu)}{V_{0}(\nu)} \right)^2 = \left( \frac{\mathcal{M}_{0}}{\mathcal{M}_l} \right) \frac{\mathcal{M}_{0}\eta_{0}}{\mathcal{M}_{\mathrm{env}}\eta_{\mathrm{env}} + \mathcal{M}_{\mathrm{core}}\eta_{\mathrm{core}}}\;,
    \label{eq:V_l_full}
\end{align}
\citep[see also][]{2014A&A...572A..11G,2017A&ARv..25....1H,2019EAS....82..189B} where $\eta_{\mathrm{env}}$ is evaluated using the scaling relation defined in Equation~\eqref{eq:belkacem+12}, $\eta_{\mathrm{core}}$ using the asymptotic expression for the core radiative damping given by Equation~\eqref{eq:eta_core} and we assume $\eta_{0} \approx \eta_{\mathrm{env}}$ This approach takes into account the state-of-the-art scaling relations for the mode damping rates. Including these damping rates provides a lower limit for the non-radial mode amplitudes (see discussion in App.~\ref{ap:3}) and it consequently provides a lower limit to the efficiency of the mixed mode mechanism.

\par Taking into consideration that the non-radial mode amplitudes can potentially be underestimated by this approach, we also computed models where we neglected the radiative damping in the core ($\eta_{\mathrm{core}} \approx 0$), and therefore used the following expression,
\begin{align}
    \left( \frac{V_l (\nu)}{V_{0} (\nu)} \right)^2 \approx \left( \frac{\mathcal{M}_{0} }{\mathcal{M}_l} \right)\;,
    \label{eq:V_l_nodamp}
\end{align}
\citep{2015A&A...579A..30B,2019EAS....82..189B} where we also assume $\mathcal{M}_{\mathrm{env}}\eta_{\mathrm{env}} \approx \mathcal{M}_{0}\eta_{0} $.
The p-dominated mode amplitudes remain similar to the first approach, whereas the g-dominated mode amplitudes are increased. This crude approximation establishes an upper limit for non-radial mode amplitudes from theoretical predictions and consequently the maximum efficiency predicted for the mixed mode mechanism. 

\par From here on, when discussing models with and without radiative damping, we will always refer to the effect of the radiative damping on the mixed mode amplitudes in the core of the star (i.e., the effect on Eq.~\ref{eq:j_dot}). These two approaches define a range within which we investigated whether the mixed modes can potentially affect the angular momentum of the star. The theoretical non-radial mode amplitudes predicted using these approaches are both well within the observational constraints from the literature \citep[e.g.,][see discussion in App.~\ref{ap:3}]{2012A&A...548A..10M,2017A&A...598A..62M}. For a more precise quantitative estimate of the effect of mixed modes on the angular momentum transport one would require measurements of absolute non-radial mode surface velocities along the RGB which is beyond the scope of the current work.

\section{Physical inputs}
\label{sec:2}

\par In this section we describe the physical inputs included in our stellar models, in particular we describe how we implemented angular momentum transport. 

\begin{figure}
   \raggedright
   \includegraphics[width=9cm]{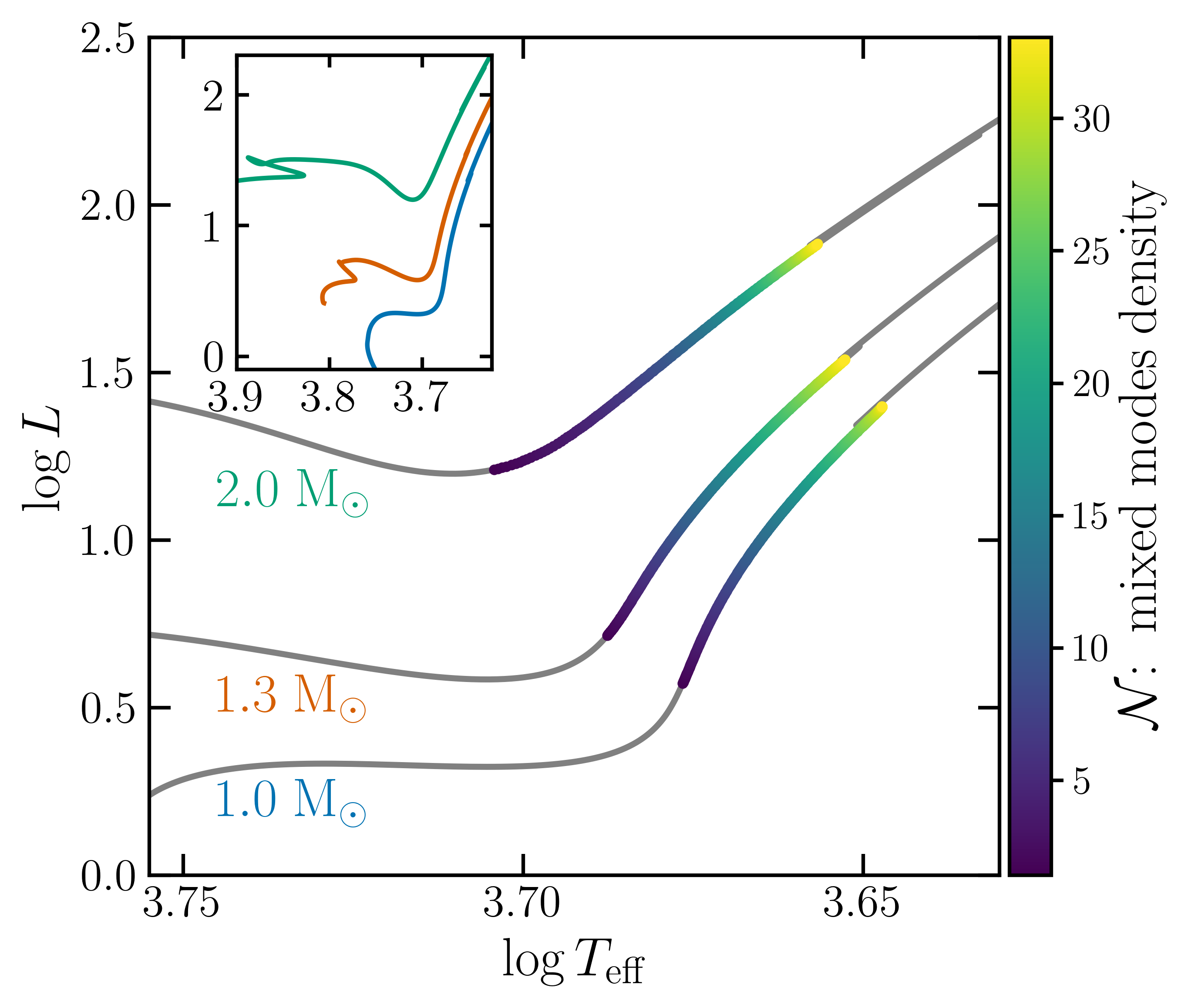}
   \caption{Hertzsprung-Russel (HR) diagram for three masses: $1.0 \;\mathrm{M}_{\odot}$ (blue line), $1.3 \;\mathrm{M}_{\odot}$ (orange line) and $2.0 \;\mathrm{M}_{\odot}$ (green line) with solar metallicity. The selected models for which we computed rotation rates are colour-coded with the correspondent mixed mode density $\left(\mathcal{N}\right)$ and range from the base of the RGB ($\mathcal{N}\sim 5$) until the RGB luminosity bump ($\mathcal{N}\sim 30$). The smaller panel shows a zoom-out of the evolutionary tracks starting from the ZAMS.}
    \label{fig:HR}
\end{figure}

\subsection{Stellar models}

\par Figure~\ref{fig:HR} shows the stellar evolution tracks for three stellar masses 1.0, 1.3 and 2.0 M$_{\odot}$ at solar metallicity computed with the release version 23.05.1 of Modules for Experiments in Stellar Astrophysics \citep[MESA,][]{paxton_modules_2011, paxton_modules_2013, paxton_modules_2015, paxton_modules_2018, paxton_modules_2019,Jermyn2023}. The standard physical inputs included in our models are detailed in Appendix~\ref{ap:1}.
\par The models for which we computed the angular momentum transport by mixed modes (marked with coloured dots in Fig.~\ref{fig:HR}) were selected to have the same range of mixed mode density, $\mathcal{N}\in [1.5-33]$. The mixed mode density can be used as a proxy for the evolution along the RGB (i.e. monotonically increasing along the RGB, as shown in Fig.~\ref{fig:HR}). This quantity is defined as the number of gravity-dominated mixed modes per radial order \citep[e.g.,][]{2018A&A...616A..24G},
\begin{align}
\label{eq:mixedmodesdensity}
    \mathcal{N} = \frac{\Delta\nu}{\Delta\Pi_1 \nu_{\mathrm{max}}^2}\;,
\end{align}
with $\Delta\nu$ the large separation, $\Delta\Pi_1$ the dipole mode period spacing and $\nu_{\mathrm{max}}$  the frequency of the maximum height in the oscillation power spectrum. 
We started computing the angular momentum transport by mixed modes at the base of the RGB, immediately after the subgiant phase where the mixed modes are not efficient at slowing down the core. The last models considered are located just before the RGB luminosity bump, since the mixed modes inertia seems to be affected by the bump.

\par We computed the mixed mode momentum flux only including mixed modes up to the angular degree $l=3$. To compute the radial and non-radial mode ($l=0,1,2,3$) eigenfrequencies, inertias and eigenfunctions we used the open source pulsation code GYRE version 7.1 \citep{2013MNRAS.435.3406T,2018MNRAS.475..879T} that solves the fourth-order system of adiabatic equations \citep[see e.g.,][]{1989nos..book.....U}. The effect of rotation on the eigenfunctions was not accounted for.

\subsection{Including rotation in stellar models}

\label{sec:rotmodels}
\par Considering the regime of low mass and evolved stars, i.e. slow rotators, we assumed that the effects on the structure of the star due to rotation can be neglected. Therefore, to incorporate rotation in our models we developed a post-processing code to solve the advection-diffusion angular momentum transport equation (combination of Eq.~\ref{eq:AM},~\ref{eq:AMshear},~\ref{eq:AMcirc} and \ref{eq:j_dot}) in the radiative zones, 
\begin{equation}
\label{eq:AM_num}
   \rho\frac{\mathrm{d} }{\mathrm{d}t} \left(r^2 \Omega\right) = \frac{1}{r^2}\frac{\partial}{\partial r} \left(\rho r^4 \nu_{\mathrm{v}} \frac{\partial \Omega}{\partial r}\right)  + \frac{1}{5 r^2}\frac{\partial}{\partial r} \left(\rho r^4 \Omega U_2 \right) + \dot{\mathcal{J}}\;.
\end{equation}

\par We follow the implementation of \cite{2013A&A...549A..74M} (embedded in the code CESTAM) where they have used the numerical relaxation scheme described by \citet{1964ApJ...139..306H, press_etal:1992}. Shear-induced turbulence and meridional circulation (first and second term on the right-hand side of Equation~\ref{eq:AM_num}) were included as described in \cite{2013A&A...549A..74M} -- following the prescriptions by \cite{1997A&A...317..749T} for the $\nu_{\mathrm{v}}$, by \cite{1998A&A...334.1000M} for the $U_2$ and by \cite{2004A&A...425..243M} for the horizontal component of turbulent diffusivity. To improve numerical stability, the mixed modes flux (last term of Eq.~\ref{eq:AM_num}) was modified to be constant in the most inner layers of the star (see full details in App.~\ref{ap:Mixed modes momentum flux}). This modification induces no significant effects on the final rotation rate and results in a negligible loss of angular momentum compared to the unmodified flux.

\par In convective zones we assume very efficient transport of angular momentum leading to solid body rotation, a common assumption in 1D stellar evolution codes \citep[e.g.,][]{palacios_rotational_2003,2013A&A...549A..74M}. The total angular momentum is conserved by including the boundary condition at the top boundary of the radiative zone:
\begin{align}
\label{eq:boundary_condition}
   \frac{\mathrm{d} }{\mathrm{d}t} \int_{m_c}^{M_{\star}} \left(r^2 \Omega\right) \;\mathrm{d}m = - \frac{4\pi}{5} \rho r^4 \Omega U_2\biggl|_{m=m_c}  + \int_{0}^{m_{\mathrm{c}}} \frac{\dot{\mathcal{J}}}{\rho} \;\mathrm{d}m\;,
\end{align}
where $m_{\mathrm{c}}$ is the mass of the base of the convection zone and $M_{\star}$ is the mass at the surface. This boundary condition ensures that the momentum flux from the boundary of the radiative region is transferred to the convective envelope.

\par At each time step our code takes as input an equilibrium structure model previously computed using MESA. We assumed an initial rotation profile with a step at the hydrogen-burning shell to start the computation (profile A in Fig.~\ref{fig:omega_evolution_diffini}). Subsequently, we computed the momentum fluxes due to the different mechanisms and solve the angular momentum transport equation. Finally, the code outputs the converged rotation profile $\Omega(r)$ which will be the input rotation profile for the next model. Hence, no feedback into the structure model is provided. In between each time step the mesh in mass coordinates is adapted automatically following the gradient method as described in \cite{1994A&A...286..121W} and \cite{weiss_garstecgarching_2008}. The code retries to solve Equation~\eqref{eq:AM_num} with smaller time steps in case the convergence criteria is not satisfied in the numerical scheme. We perform a linear interpolation of the mixed mode momentum flux ($\dot{\mathcal{J}}$) for the smaller time steps, since this quantity evolves almost monotonically over time.

\par The MESA and GYRE inlists, and the output files from the post-processing angular momentum transport code used in this work are publicly available via Zenodo (DOI: \href{https://zenodo.org/records/15553726}{10.5281/zenodo.15553726}).

\section{Timescales associated with angular momentum transport}
\label{sec:timescales}

\par In this section we compare the timescales associated with angular momentum transport by several mechanisms to understand the effect of the mixed modes on the rotation profiles. Here we focus on the 1.3 M$_{\odot}$ model, see Appendix~\ref{ap:Timescales} for a discussion on the 1 M$_{\odot}$ and 2 M$_{\odot}$ models. Therefore, as in \cite{2006A&A...453..261P} and \cite{2015A&A...579A..30B} we define:
\begin{itemize}
    \item $\tau_{\mathrm{modes}} \sim \left|\rho r^2\Omega /\dot{\mathcal{J}} \right|$ as the timescale associated with angular momentum transport by mixed modes;
    \item $\tau_{\mathrm{contr}}\sim \left|r/\dot{r}\right|$ as the timescale associated with the contraction of the star, where $\dot{r} = \mathrm{d}r/\mathrm{d}t$;
    \item $\tau_{\mathrm{shear}}\sim \left| \left(\Delta r\right)^2 \Omega/\left(\nu_{\mathrm{v}}\Delta\Omega\right)\right|$ as the characteristic timescale associated with angular momentum transport by shear-induced turbulence;
    \item $\tau_{\mathrm{circ}}\sim \left|\Delta r/U_2 \right|$ as the characteristic timescale associated with angular momentum transport by meridional circulation.
\end{itemize}

\begin{figure}
    \centering
    \includegraphics[width=0.5\textwidth]{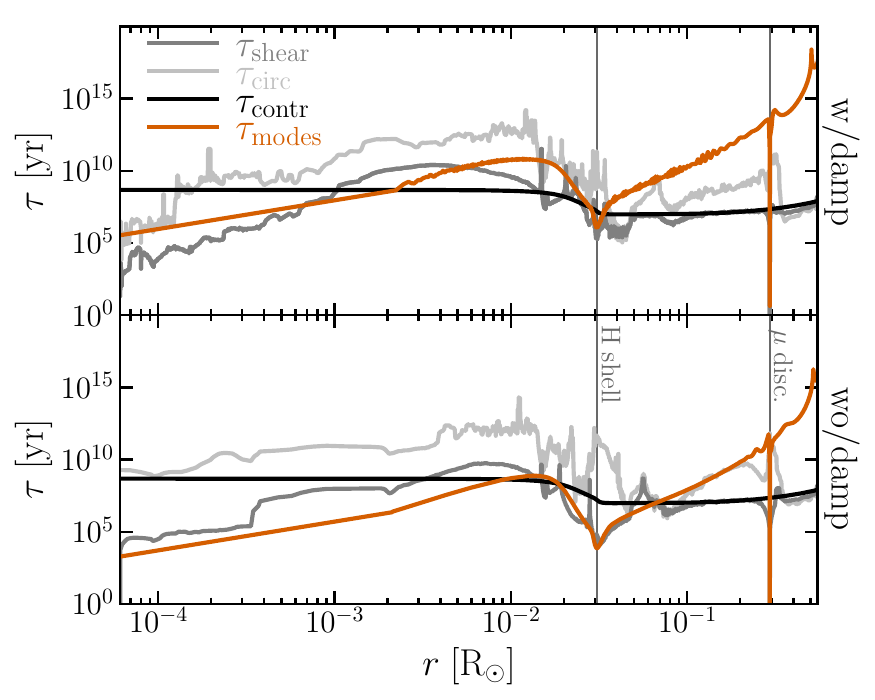}
    \caption{Timescale (logarithmic scale) associated with the different mechanisms responsible for angular momentum transport as a function of stellar radius (logarithmic scale) for a model with mass of $1.3\;\mathrm{M}_{\odot}$ and $\mathcal{N}\sim30$ (close to RGB bump). The line colour indicates the angular momentum transport mechanism. The top (bottom) panel corresponds to models with (without) radiative damping, respectively. The grey vertical lines show the position of the hydrogen-burning shell (labelled as H-shell) and of the chemical discontinuity (labelled as $\mu$-disc.), respectively.}
    \label{fig:timescales}

    \centering
    \includegraphics[width=0.5\textwidth]{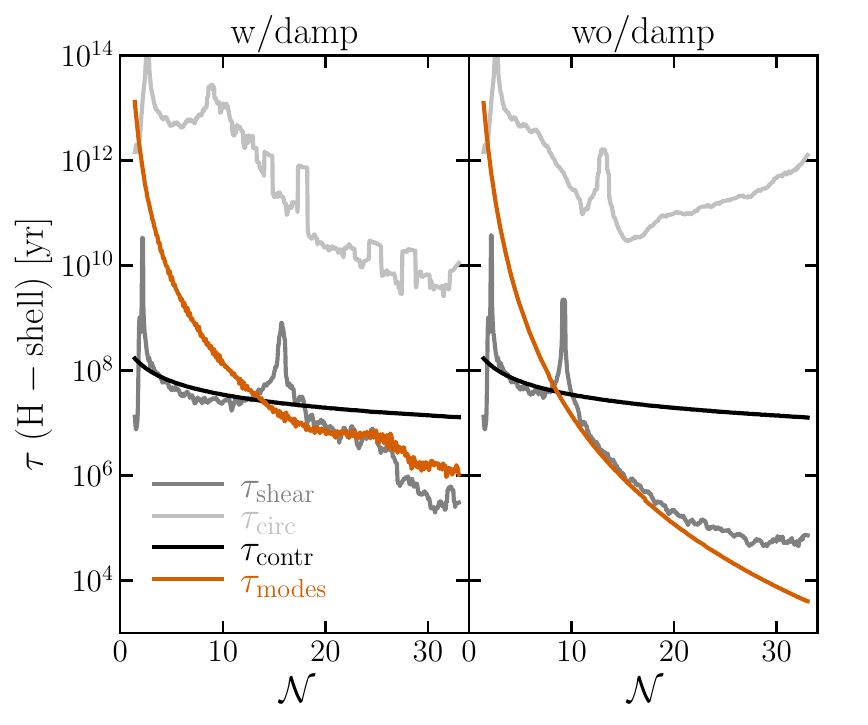}
    \caption{Timescale (logarithmic scale) evaluated at the hydrogen-burning shell as a function of mixed modes density for the $1.3\;\mathrm{M}_{\odot}$ stellar evolution track. The line colours have the same meaning as in Fig.~\ref{fig:timescales}. The left (right) panel corresponds to models with (without) radiative damping, respectively.}
    \label{fig:timescales_}
\end{figure}

\subsection{Timescales within the radiative region}

\par In Figure~\ref{fig:timescales} we show the timescales associated with the mechanisms responsible for angular momentum transport at a particular time in evolution (just below the RGB bump at $\mathcal{N}=30$). The mechanism with the smallest timescale naturally dominates the angular momentum transport in a particular region of the star. 

\par The mixed modes are most efficient around the hydrogen-burning shell, in the chemical discontinuity produced by the first dredge-up, and in the very centre of the star \citep[see orange line in Fig.~\ref{fig:timescales} and also][]{2015A&A...579A..30B}. We note that in the latter the percentage of angular momentum extracted compared with the total angular momentum of the star is very small since it covers a very small fraction of mass of the star. In these regions where the mixed modes dominate, the mixed modes counteract the spin up due to contraction (black line in Fig.~\ref{fig:timescales}) consequently slowing down the core. In particular, for the models neglecting radiative damping in the core, the mixed modes extract angular momentum faster than the shear-induced turbulence (medium grey line in Fig.~\ref{fig:timescales}) can diffuse the angular momentum (redistribute to adjacent cells) resulting in an extremely localised extraction of angular momentum. Due to the significant localised extraction of angular momentum and the very small timescales around those regions, several numerical difficulties arise when solving the angular momentum transport equation (Eq.~\ref{eq:AM_num}). For that purpose we included an automated adaptive mesh and adaptive time step, as mentioned in Section~\ref{sec:rotmodels}, to increase the mesh resolution in those regions and to perform successive iterations with smaller time steps.

\par In the regions where the mixed modes are not efficient, in particular, in the shells just below the hydrogen-burning shell, the contraction of the star dominates and consequently spins up those shells. In the remaining shells, the shear-induced turbulence is able to diffuse and redistribute angular momentum extracted by mixed modes. Hence, the extraction of angular momentum by mixed modes will be less localised. The meridional circulation is mostly inefficient at redistributing angular momentum when compared with any other mechanism despite an increase in its efficiency in shells near the hydrogen-burning (light grey line in Fig.~\ref{fig:timescales}).

\par We also show that the timescale $\tau_{\mathrm{modes}}$ is roughly two orders of magnitude higher for models including radiative damping than for models without mode damping, for the same point in evolution. This shows that the inclusion of radiative damping significantly reduces the efficiency of the mixed mode mechanism. 
It is also important to note that the shear-induced turbulence is able to redistribute the localised spin down due to the mixed modes in the models including damping (medium grey line overlaps with orange line in the top panel in Fig.~\ref{fig:timescales}), whereas in the models neglecting radiative losses, the effect of the mixed modes remains very localised (medium grey line is above the orange line in the bottom panel in Fig.~\ref{fig:timescales}).

\subsection{Timescales evolution along the RGB}

\par In Figure~\ref{fig:timescales_} we evaluate the timescales associated with angular momentum transport at a particular location inside the star, that is at the hydrogen-burning shell. This allows us to understand how the timescales evolve along the RGB and how they affect this particularly important region in the star.

\par As discussed in \cite{2015A&A...579A..31B}, we see that the efficiency of the mixed mode mechanism increases along evolution on the RGB (see orange lines in Fig.~\ref{fig:timescales_}). This was already expected since the mixed modes density and amplitude increase along evolution, only becoming relevant for the angular momentum transport later on the RGB. From the left and right panel in Figure~\ref{fig:timescales_}, we infer that the mixed mode mechanism becomes efficient at extracting angular momentum around $\mathcal{N}=14$ for the models with mode damping, and around $\mathcal{N}=8$ for the models without mode damping. Comparing the medium grey and orange lines in the right panel in Figure~\ref{fig:timescales_}, it is evident that in the models without mode damping, i.e. with higher mixed mode amplitudes, the mixed mode mechanism is more efficient at extracting angular momentum than in the models with mode damping (see left panel in Fig.~\ref{fig:timescales_}). 
Once again this reveals that the efficiency of this mechanism is strongly correlated with the mode amplitudes. Overall, we see that including radiative damping significantly affects the efficiency of the mixed mode mechanism, or in other words, the spin down of red giant's cores due to mixed modes is delayed in models accounting for mode damping.

\par Early on the RGB, the contraction of the core and expansion of the envelope mostly dominate the evolution of the rotation profile (see black line in Fig.~\ref{fig:timescales_}). 
The shear-induced turbulence is also an efficient mechanism in our models along the RGB (medium grey line in Fig.~\ref{fig:timescales_}) and its efficiency is affected by the appearance of steep gradients in the rotation profile. In particular, this timescale spikes just after the mixed modes become efficient (see medium grey lines in Fig~\ref{fig:timescales_}) due to a sign reversal on the rotation rate gradient. For $\tau_{\mathrm{modes}} > \tau_{\mathrm{contr}}$, the rotation rate gradient  is positive around the hydrogen-burning shell. For $\tau_{\mathrm{modes}} < \tau_{\mathrm{contr}}$, the rotation rate gradient becomes negative due to the localised spin down around the hydrogen-burning shell caused by the mixed modes. In contrast, the timescale of the meridional circulation is always several orders of magnitude higher than the other mechanisms (light grey line in Fig.~\ref{fig:timescales_}), making it not efficient at redistributing angular momentum along the RGB. 

\begin{figure*}
   \centering
   \includegraphics[width=1.\textwidth]{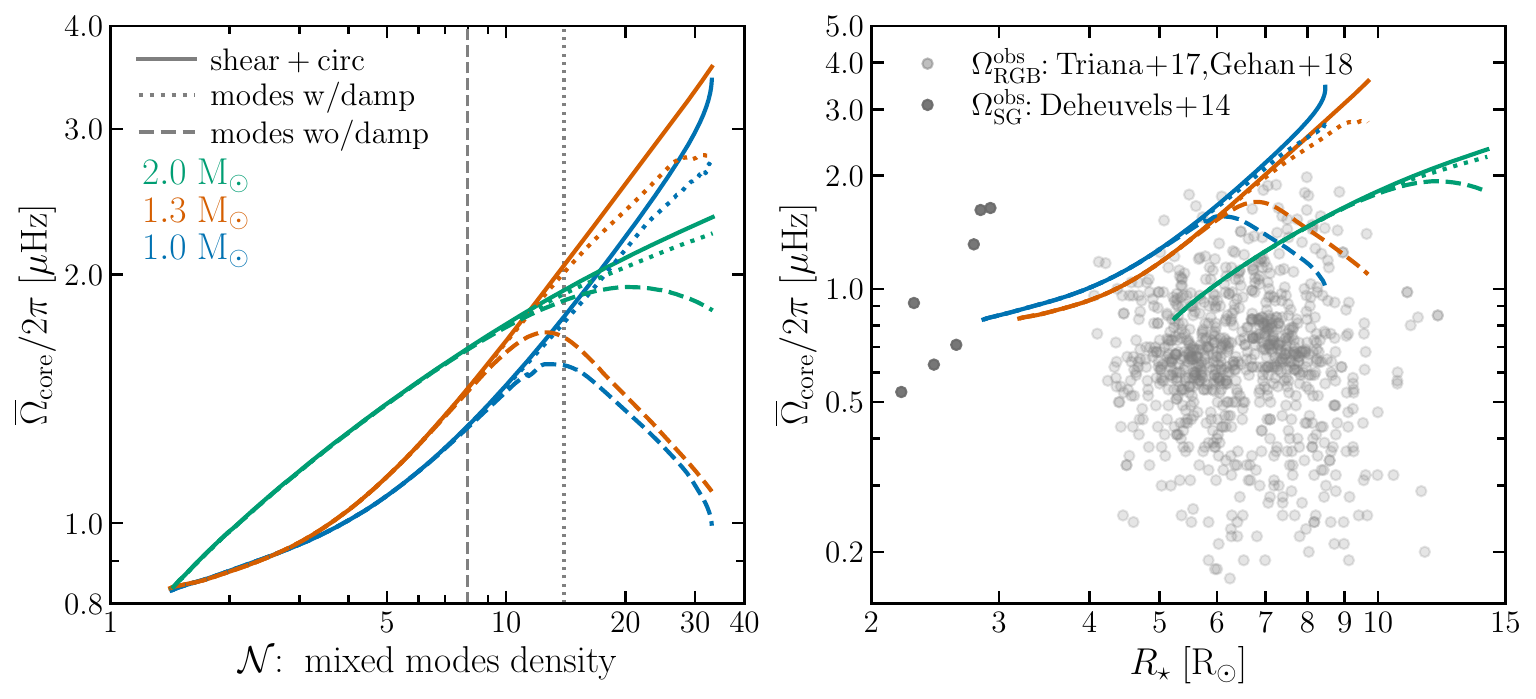}
   \caption{Core rotation rates along evolution. Left panel: Average core rotation rate (logarithmic scale) as a function of mixed mode density. Right panel: Average core rotation rate (logarithmic scale) as a function of stellar radius (logarithmic scale). The different colours represent evolutionary tracks with different stellar masses and the different line-styles indicate the angular momentum transport mechanisms included as per the legend. The dashed (dotted) vertical grey lines in the left panel indicate the values $\mathcal{N}=8$ ($\mathcal{N}=14$), respectively. The grey circles in the right panel indicate the observed core rotation rates from \cite{2014A&A...564A..27D,2017A&A...602A..62T,2018A&A...616A..24G}.}
    \label{fig:omega_evolution_}
\end{figure*}

\section{Quantitative estimate of spin down of red giants cores by mixed modes}
\label{sec:Quantitative estimate of spin down of red giants cores by mixed modes}

\par In this section we present the core rotation rates obtained when solving the angular momentum transport Equation~\eqref{eq:AM_num}. We show the comparison between including angular momentum transport by mixed modes with amplitudes estimated with and without radiative mode damping.
We subsequently compare our results with observations and show the effect of the mixed mode mechanism on the rotation profile.

\subsection{Evolution of rotation rates}

\par To investigate the effect of the mixed modes on the rotation rate along evolution and to compare with asteroseismic observations, we computed the average core rotation rate sensed by a gravity mode. 
The average rotation rate is then defined as
\begin{align}
    \overline{\Omega} = \frac{1}{\tau_{\mathrm{travel}}} \int \Omega(r)\;\mathrm{d}\tau_{\mathrm{travel}}\;,
    \label{eq:average_omega}
\end{align}
\citep{2013A&A...549A..75G} where $\tau_{\mathrm{travel}}$ is the time it takes for a wave packet to travel back and forth in the core, and is given by
\begin{align}
    \tau_{\mathrm{travel}} \sim \frac{2}{\sigma_R} \int k_r\;\mathrm{d}r \approx \frac{2\sqrt{l(l+1)}}{\sigma_R^2} \int \left(N^2 - \sigma_R^2\right)^{1/2} \frac{\mathrm{d}r}{r} \;,
    \label{eq:tau_travel}
\end{align}
\citep{1989nos..book.....U,2013A&A...549A..75G} where we assumed $\sigma_R^2~\ll~S_l^2$, which is a valid assumption within the g-mode cavity. Combining Equation~\eqref{eq:average_omega} and Equation~\eqref{eq:tau_travel}, the average core $\overline{\Omega}_{\mathrm{core}}$ rotation rate can be written as follows
\begin{align}
    &\overline{\Omega}_{\mathrm{core}} = \frac{1}{\int_{\mathrm{core}} \left(N^2 - \sigma_R^2\right)^{1/2} \frac{\mathrm{d}r}{r}} \int_{\mathrm{core}}\Omega(r) \left(N^2 - \sigma_R^2\right)^{1/2} \frac{\mathrm{d}r}{r} \;,
    \label{eq:final_average_coreomega}
\end{align}
where the bounds of the integrals correspond to the bounds of the gravity mode cavity defined by $\sigma_R < N$. In the core, we estimated the travel time of a gravity mode with $\sigma_R~=~2\pi \nu_{\mathrm{max}}$ \citep[see e.g.,][]{2019MNRAS.485.3661F}. For the entire envelope we adopted a constant value since we enforced solid body rotation in our computation, $\overline{\Omega}_{\mathrm{env}} = \Omega(r_{\mathrm{c}})$, where $r_{\mathrm{c}}$ is the radius of the base of the convection zone. 

\par
\begin{figure*}
    \raggedright
    \includegraphics[width=1.0\textwidth]{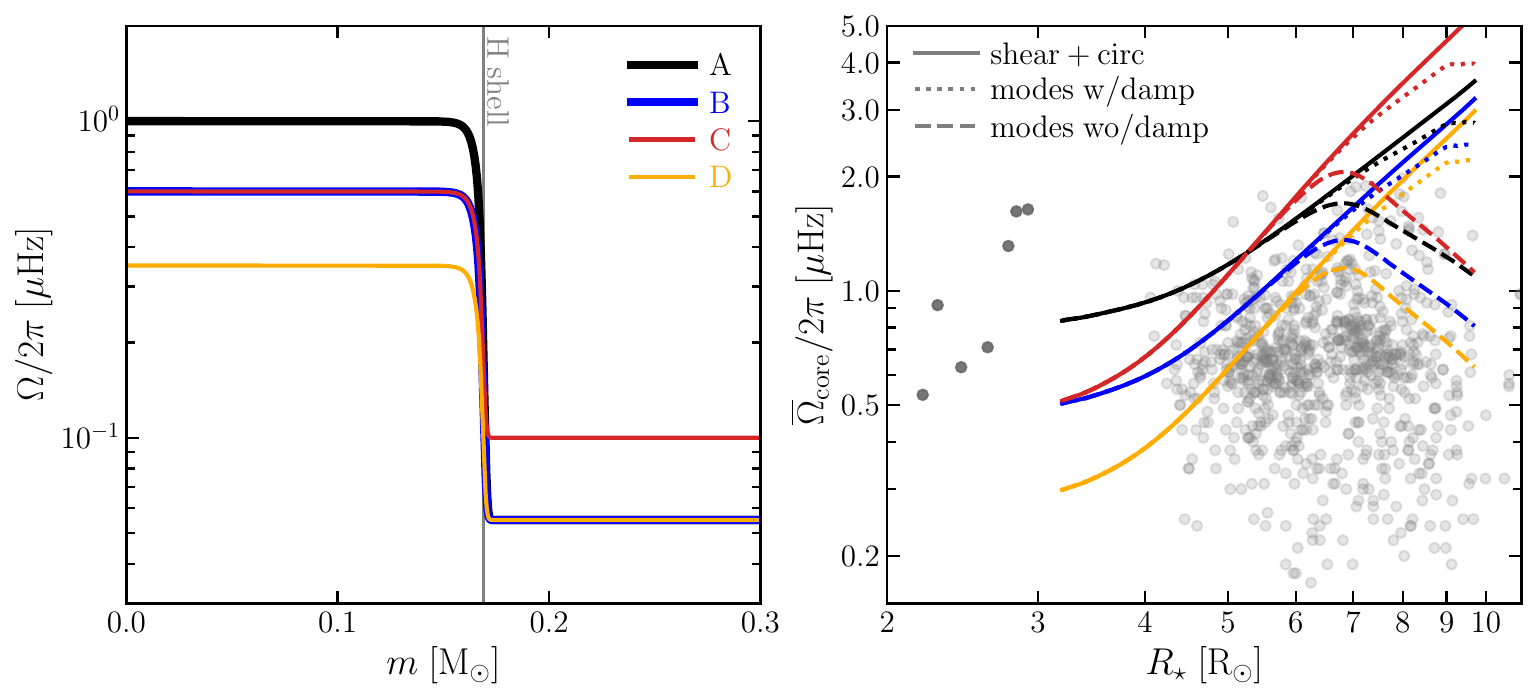}
    \caption{Initial rotation profiles and rotation rates along evolution. Left panel: Rotation profile (logarithmic scale) as a function of the fractional mass. The different colours represent models computed with the different initial rotation profiles listed in Table~\ref{tab:rot}. Right panel: Average core rotation rate (logarithmic scale) as a function of stellar radius (logarithmic scale) for the $1.3\;\mathrm{M}_{\odot}$ stellar evolution track.  The line-style and the grey circles have the same meaning as in Fig.~\ref{fig:omega_evolution_}.}
     \label{fig:omega_evolution_diffini}
 \end{figure*}

 In Figure~\ref{fig:omega_evolution_} we show the evolution of the average core rotation rate when including different angular momentum transport mechanisms. We omit the envelope rotation rates for the current discussion. Our computations were initialized
with a rotation profile with a step at the hydrogen-burning shell (profile A in Fig.~\ref{fig:omega_evolution_diffini}).

\begin{itemize}
    \item The models hereafter referred to as '\textit{shear\;+\;circ}', and indicated with solid lines, were computed accounting for redistribution of angular momentum by shear-induced turbulence and by the meridional circulation as described in Section~\ref{sec:rotmodels}. These models are shown for comparison purposes. The final rotation rate of these models is mostly a result of the contraction of the core since the meridional circulation and shear-induced turbulence are not efficient at redistributing angular momentum for this particular stage of evolution \citep{2012A&A...544L...4E,2013A&A...549A..74M,2013A&A...555A..54C}. Therefore, we see a spin up of the core as the star evolves. The tracks with different masses spin up at different rates due to their inherent different radius despite being initialized with the same rotation rate. \\
    
    \item The models hereafter referred to as '\textit{modes\;w/\;damp}', and indicated with dotted lines, were computed accounting for the mixed mode mechanism including the effect of radiative damping on the mode amplitudes, in addition to the contributions from the shear-induced turbulence and the meridional circulation. The low mixed mode amplitudes (due to high damping rates, see App.~\ref{ap:3} for details) included in these models limit the effect of the mixed modes on the angular momentum transport. The average core rotation rate for these models mostly reflects the contraction of the core up to $\mathcal{N}=14$. After that point the core spins up at a lower rate due to the extraction of angular momentum by the mixed modes.\\
    
    \item Lastly, the models hereafter referred to as '\textit{modes\;wo/\;damp}', and indicated with dashed lines, include the mixed mode mechanism neglecting the effect of radiative damping on the mode amplitudes, in addition to the contributions from the shear-induced turbulence and the meridional circulation. The higher mixed mode amplitudes included in these models enhance the effect of the mixed modes on the angular momentum transport. The average rotation rate for these models also reflects the contraction of the core however only up to $\mathcal{N}=8$. After that point the core slows down significantly. 
\end{itemize}

\begin{figure*}[ht]
    \raggedright
    \includegraphics[width=\textwidth]{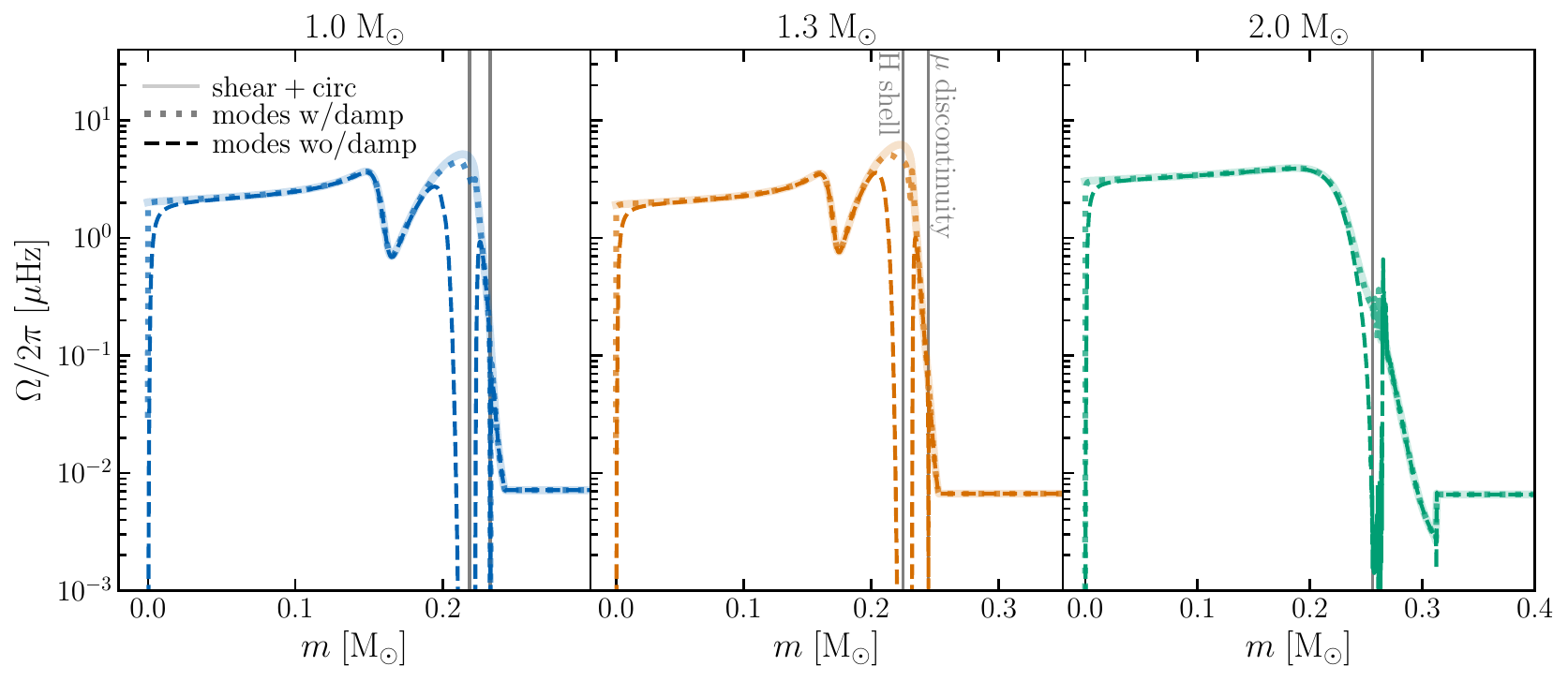}
    \caption{Rotation rate (logarithmic scale) as a function of fractional stellar mass for models with mixed modes density of $\mathcal{N}=30$. The line colours and line-styles have the same meaning as in Fig.~\ref{fig:omega_evolution_}. The vertical grey lines show the position of the hydrogen-burning shell and of the chemical discontinuity, respectively.}
    \label{fig:rotprofile}
\end{figure*}

\par In the models accounting for the extraction of angular momentum by mixed modes the core clearly rotates slower than the models only including meridional circulation and shear-induced turbulence (left panel in Fig.~\ref{fig:omega_evolution_}). It is also evident that the mixed modes are not efficient at counteracting the spin up of the core due to contraction before reaching $\mathcal{N}=8$. This suggests that another mechanism of angular momentum transport needs to be included to slow down the core of the star during the subgiant phase (e.g., magnetic fields or IGWs). Once again we see that the point in evolution where the mixed mode mechanism starts to slow down the core depends on the amount of damping of the mixed mode amplitudes. Likewise, the rate at which the core slows down depends on the amount of mode damping: the non-damped models spin down much faster than the damped models.

\par Overall our results show that the mixed mode mechanism is efficient at slowing down the core if the radiative damping is neglected. When accounting for the radiative damping predicted from asymptotic theory, the mixed mode mechanism is however not sufficient to slow down the core and other mechanisms of angular momentum transport have to be considered. Nonetheless, considering that the mixed modes amplitudes including radiative damping may be underestimated (see discussion in App.~\ref{ap:3}) we hypothesize that the actual spin down of the core due to the mixed modes would lie between the dashed and dotted tracks (i.e. core rotation rates neglecting or accounting for radiative losses on the mixed mode amplitudes). To confirm this idea one would require precise measurements of the gravity-dominated mixed modes amplitudes for angular degrees $l=2$ and $l=3$ or even higher, which is not possible due to their low amplitudes and due to cancellation effects at the surface of the star. Hereafter, we consider the effect of the mixed mode mechanism on the rotation profile to be within the range in average rotation rate defined between the dashed and dotted lines.

\par As shown by \cite{2018A&A...616A..24G} the dependence of observed rotation rates on the stellar mass, within $1-2\;\mathrm{M}_{\odot}$, is quite small. In the track with $1.3\;\mathrm{M}_{\odot}$ the core spins up slightly more than for the $1.0\;\mathrm{M}_{\odot}$ track, solely due to the contraction of the inherently larger core. When including the mixed mode mechanism without mode damping both tracks spin down to similar values of the core rotation, suggesting that the spin down due to the mixed modes is independent of the stellar mass (see blue and orange lines in left panel in Fig.~\ref{fig:omega_evolution_}). The difference in both tracks increases however when accounting for mode damping in the mixed modes, since for both tracks the evolution is still partially dominated by the contracting core. Nonetheless, the point in the evolution where the mixed modes start to slow down the core seems to be independent of stellar mass as well. For the $2.0\;\mathrm{M}_{\odot}$ model, as discussed in Appendix~\ref{ap:Timescales}, the contraction of the core is more efficient than for lower masses, particularly early in the RGB (see green lines in left panel in Fig.~\ref{fig:omega_evolution_}). Therefore, the effect of the mixed modes on the rotation rate is less significant and delayed in evolution than for other masses.

\subsection{Comparison with observations}

\par In Figure~\ref{fig:omega_evolution_} (right panel) we show the evolutionary tracks with asteroseismic data as a function of stellar radius for comparison purposes. The observed subgiant and red giant core rotation rates were compiled by \cite{2019ARA&A..57...35A} using the data from \cite{2014A&A...564A..27D}, \cite{2017A&A...602A..62T} and \cite{2018A&A...616A..24G}. Typical errors for the mixed-mode splittings, hence for the rotation rates, are around 1\% \citep{2019ARA&A..57...35A}. 

\par We find that the rotation rates computed with the mixed mode mechanism without mode damping overlap with the observed core rotation rates, particularly for the $1.0\;\mathrm{M}_{\odot}$ and $1.3\;\mathrm{M}_{\odot}$ (Fig.~\ref{fig:omega_evolution_} right panel). However, when accounting for radiative damping in the core the computed rotation rates are higher than the observed ones. The range defined by these two tracks, which we argue is where a track computed with realistic mixed mode amplitudes would lie, overlaps only partially with observations. This suggests that even though the mixed modes are able to slow down the core of the star, they are not able to reach the observed values for the core rotation. In Section~\ref{sec:Different initial rotation profiles} we investigate if the final rotation rates obtained when including the mixed mode mechanism depend on the initial rotation rates or on the initial rotation profiles. Other additional factors such as magnetic braking, mass loss and the effect on the centrifugal force on the stellar structure could potentially explain some of the spread in the observed rotation rates. We do not take any of those factors into account in the present work.

\par Since in our models the efficiency of the mixed modes depends on whether we take into account radiative damping in the core or not, it may be possible to constrain the mixed mode amplitudes (or/and constrain the damping rates) by matching the computed rotation rates with observations. This assumes that our current theory for the angular momentum transport is correct, which at the moment we are not able to prove. Another more plausible conclusion is that an additional mechanism of angular momentum transport is required to spin down the core of the star in this stage of evolution. Further investigation on the interaction of the mixed mode mechanism with an additional mechanism is crucial in order to understand the full extent of the mixed mode mechanism -- this is foreseen in a future work.

\subsection{Effect on the rotation profiles}
\label{sec:Effect in the rotation profiles}

\par In this section we now discuss the effect of the mixed mode mechanism on the rotation profiles of individual models. 

\par In Figure~\ref{fig:rotprofile} we show rotation profiles for models with mixed mode density of 30 (i.e. close to the RGB bump) with different stellar masses. The rotation profiles for $1.0\;\mathrm{M}_{\odot}$ and $1.3\;\mathrm{M}_{\odot}$ show two bumps just below the hydrogen-burning shell. The outer bump is the result of the outer shells in the core contracting faster than the innermost shells. The inner bump on the other hand is due to the step in the initial rotation profile. 
In the $2.0\;\mathrm{M}_{\odot}$, there are no bumps in the rotation profile. This is a consequence of the nearly uniform contraction in the core and of the step on the initial rotation profile almost coinciding with the stationary shell in the star (i.e., the location in the star where $\mathrm{d} r/\mathrm{d} t = 0$) which is where the differential rotation develops in our models. Also, for the $2.0\;\mathrm{M}_{\odot}$, the more efficient contraction yields higher core rotation rates than for lower masses. See Appendix~\ref{ap:MESA} for a comparison between our models without mixed modes and MESA rotating models.

\par In the models where we accounted for radiative damping, the effect of the mixed modes is visible by a small reduction of the rotation rate in the hydrogen-burning shell and by a steep decrease of the rotation rate in the innermost layer of the star (dotted lines in Fig.~\ref{fig:rotprofile}).  
Whereas in the models without radiative damping we see more significant changes in the rotation profile -- there is a significant decrease of rotation rate around the hydrogen-burning shell in addition to the steep decrease in the innermost layer of the core (dashed lines in Fig.~\ref{fig:rotprofile}). In both these regions the rotation rate reaches very low and negligible values compatible with zero. This suggests that those shells stop rotating at a particular moment in the evolution, which seems an unlikely scenario. 

\par Recent efforts to disentangle the rotation profile from observations point towards a uniform rotating helium core with a shear layer between the rapidly rotating core and slowly rotating envelope that lies within the hydrogen-burning shell \citep{2017MNRAS.464L..16K,2018ApJ...862....9D}. Despite the non-uniform rotating helium core in our models, the localised spin down in the hydrogen-burning shell due to the mixed modes could be compatible with the steep layer in these studies. More detailed information on the rotation profile in radiative regions will certainly provide stronger constraints on the physical processes dominating the angular momentum transport in red giants.

\par From the timescales in Figure~\ref{fig:timescales} and~\ref{fig:timescales_} we could already infer that both the meridional currents and the shear-induced turbulence would not be enough to redistribute the localised spin down due to the mixed modes \citep[contrary to what was suggested by][]{2015A&A...579A..31B}. From Figure~\ref{fig:rotprofile}, we can certainly confirm that the mixed modes effect on the rotation profile remains localised even when coupled with meridional circulation and shear-induced turbulence coefficients. 
It is reasonable to assume that the interaction of the mixed mode mechanism with another efficient mechanism of angular momentum transport (i.e., with a higher viscosity coefficient) would be sufficient to redistribute the localised spin down due to the mixed modes and further increase the core spin down. One candidate would be, for example, magnetohydrodynamic instabilities such as the Tayler-Spruit dynamo \citep{spruit_dynamo_2002} or the most recent formulations \citep{2019MNRAS.485.3661F,2022A&A...664L..16E}. In future work we aim to provide a quantitative estimate to verify this hypothesis, for now, to test this idea we performed a simplified test using a constant viscosity coefficient. More details regarding this test are described in Section~\ref{sec:Including an additional diffusion coefficient}.

\section{Testing the efficiency of the mixed mode mechanism with different physics}
\label{sec:Testing the efficiency of the mixed mode mechanism with different physics}

\begin{table}
\caption{\label{tab:rot} Values of core and envelope rotations rates in our initial rotation profiles.}
\centering
\begin{tabular}{lcc}
\hline\hline
Profile & $\Omega_{\mathrm{core}}/2\pi$ ($\mu$Hz) & $\Omega_{\mathrm{env}}/2\pi$ ($\mu$Hz) \\
\hline
A           &1.00     &0.05 \\
B           &0.60    &0.05\\
C           &0.60   &0.10\\
D           &0.35      &0.05\\
\hline
\end{tabular}
\tablefoot{Profiles A, B and C have core and envelope rotation rates compatible with observations of subgiant stars \citep{2014A&A...564A..27D}. Profile D has a core rotation rate lower than observations.}
\end{table}

\par In the previous section, our computations including the mixed mode mechanism for angular momentum transport resulted in final rotation rates that only overlapped with the higher range of observations (relatively fast rotating cores). In this section we aim to understand if by modifying some of the physical parameters in our models we can explain the full range of observed rotation rates. We started by testing different initial rotation rates in Section~\ref{sec:Different initial rotation profiles}. Then, in Section~\ref{sec:Including an additional diffusion coefficient}, we included an additional viscosity when solving the angular momentum equation to simulate a potential missing diffusion mechanism.

\subsection{Dependence on the initial rotation profile}
\label{sec:Different initial rotation profiles}

\par In this section we investigate the dependence of the rotation rates computed with the mixed mode mechanism on the initial rotation rates. All initial rotation profiles have a step in the middle of the hydrogen-burning shell. The width of the transition region corresponds to the width of the hydrogen-burning shell. The first three rotation profiles (A, B and C) are consistent with observations of subgiant stars \citep{2014A&A...564A..27D}. The following initial rotation profiles were initiated at the beginning of our computation, at the end of the subgiant phase (see left panel in Fig.~\ref{fig:omega_evolution_diffini}):

\begin{itemize}
    \item \textit{Profile~A}: this rotation profile was used in the previous sections to initialize our computations, and, imposes the highest degree of radial differential rotation amidst our profiles, with the highest core rotation rates.
    \item \textit{Profile~B}: this rotation profile has a slower core rotation and the same envelope rotation rate as \textit{Profile~A}. 
    \item \textit{Profile~C}: in this rotation profile the degree of differential rotation included is smaller (slower core rotation and faster envelope rotation). This rotation profile would be achieved with a more efficient transport of angular momentum from the core to the outer layers during the subgiant phase.
    \item \textit{Profile~D}: this is the rotation profile with the lowest core rotation rate, which is no longer consistent with the observations of subgiant stars. This profile was included as a test case. The envelope rotation rate is the same as in \textit{Profile~A}.
\end{itemize}

\par In the right panel in Figure~\ref{fig:omega_evolution_diffini} we show that the point in the evolution where the mixed mode mechanism becomes efficient seems to be independent of the initial rotation profile. Likewise, the range in between models with damping and without also seems independent of the initial core rotation rates (profiles A, B and D). However, there seems to be some dependence on envelope rotation rate (profile C).

\par The models where we only include angular momentum transport by meridional circulation and shear-induced turbulence spin up the core to rotation rates outside the observational window despite the low initial core rotation rates (right panel in Fig.~\ref{fig:omega_evolution_diffini}). The models which include angular momentum transport by mixed modes accounting for radiative damping in the core also predict final rotation rates higher than observations. This was expected since the contraction of the core dominates early in the evolution and the mixed modes spin down later in the evolution is not enough to slow down the core. On the other hand, when neglecting the radiative damping in the core we show that including slower initial core rotation in our models leads to lower final rotation rates in agreement with observations (except for the profile C). This highlights the importance of uncovering the unknown mechanism responsible to slow down the core at the end of the subgiant phase (which in this work we are simulating by enforcing different initial rotation rates). This unknown mechanism will most likely play an important role in explaining the spread we observe in rotation rates along the RGB. 

\par The models computed with different initial rotation profiles and with the mixed mode mechanism certainly produce a wider spread in the final core rotation rates than models with different stellar masses (by comparing the right panels of Fig.~\ref{fig:omega_evolution_} and Fig.~\ref{fig:omega_evolution_diffini}). 
Nonetheless, the wider spread in rotation rates from these models is not able to explain the lowest values of core rotation rates from observations, even when considering an unreasonable low initial core rotation (i.e. profile D). This re-enforces the conclusion that the interaction with other mechanisms of angular momentum transport has to be considered.

\subsection{Including an additional viscosity coefficient}
\label{sec:Including an additional diffusion coefficient}

\begin{figure}
   \raggedright
   \includegraphics[width=0.5\textwidth]{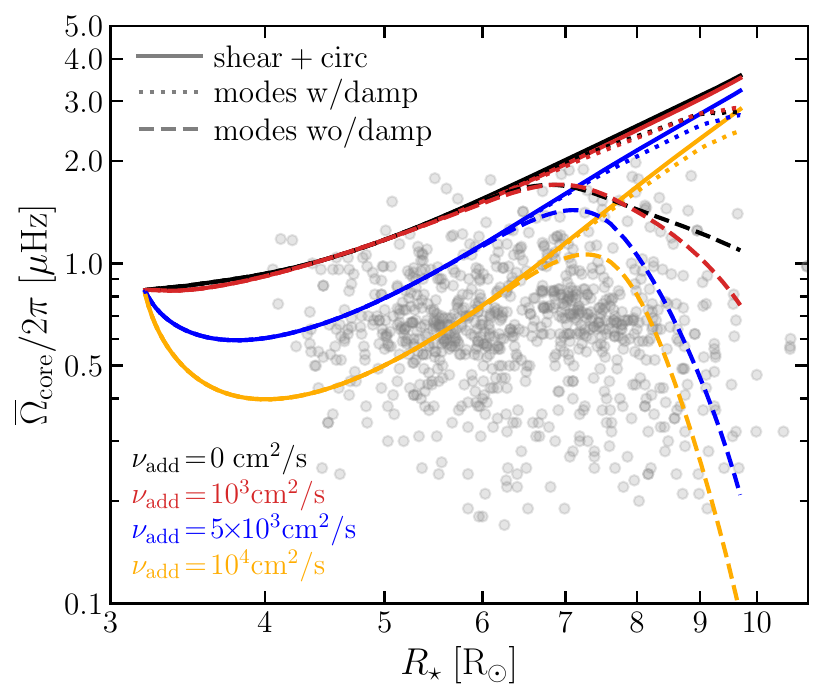}
   \caption{Average core rotation rate (logarithmic scale) as a function of the stellar radius (logarithmic scale) for the $1.3\;\mathrm{M}_{\odot}$ stellar evolution track. The different colours show models computed with different values for the additional viscosity term in Eq.~\eqref{eq:AM__}. The line-style and the grey circles have the same meaning as in Fig.~\ref{fig:omega_evolution_}.}
    \label{fig:omega_evolution_diffnu}
\end{figure}

\begin{figure*}
    \raggedright
    \includegraphics[width=\textwidth]{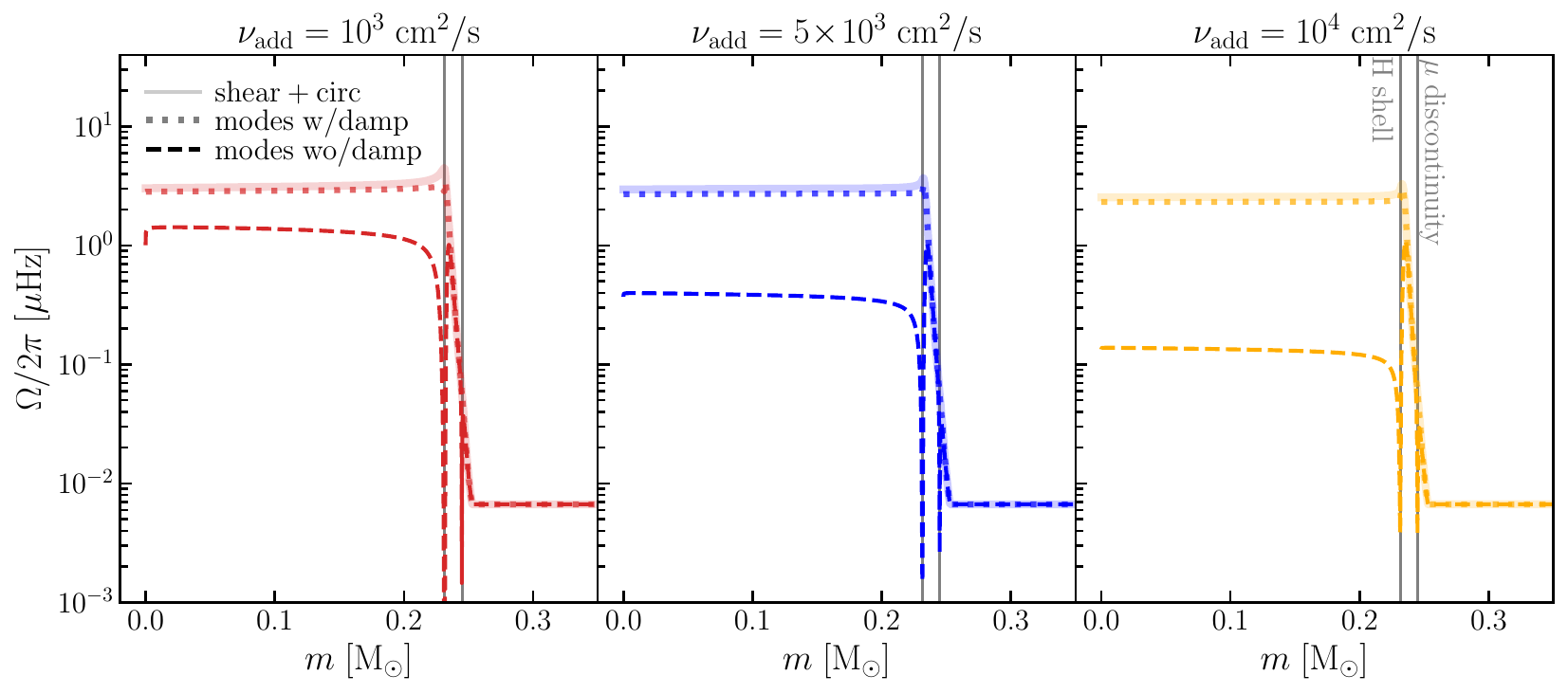}
    \caption{Rotation rate (logarithmic scale) as a function of fractional stellar mass for models with mixed modes density of 30 and $1.3\;\mathrm{M}_{\odot}$. The three panels show models computed with different values for the additional viscosity term in Eq.~\eqref{eq:AM__}. The colours represent the same angular momentum transport mechanisms as in Fig.~\ref{fig:rotprofile}.}
    \label{fig:rotprofile_diffnu}
\end{figure*}

\par In this section we investigate the hypothesis that an additional mechanism of angular momentum transport would be able to redistribute the localised spin down due to the mixed modes, therefore increasing the efficiency of angular momentum transported, and consequently slowing down the core of RGB stars to match observations.
\par We adopted a similar approach as in \citet{2017A&A...599A..18E}. They assumed the physical nature of the missing mechanism for angular momentum transport in the RGB was diffusive, and included an additional constant viscosity term in the equation of angular momentum transport (Eq.~\ref{eq:AM_num}) to simulate this mechanism. The goal is not to uncover the physics within this additional unknown mechanism, but rather to understand the effect of such an efficient mechanism on the angular momentum redistribution. For this purpose, we make the simple assumption of a constant viscosity in radius and in time. In our work, the angular momentum equation in radiative zones (Eq.~\ref{eq:AM_num}) becomes the following, 
\begin{equation}
\label{eq:AM__}
   \rho\frac{\mathrm{d} (r^2 \Omega)}{\mathrm{d}t}  = \frac{1}{r^2}\frac{\partial}{\partial r} \left(\rho r^4 \left(\nu_{\mathrm{v}}+\nu_{\mathrm{add}}\right) \frac{\partial \Omega}{\partial r}\right)  + \frac{1}{5 r^2}\frac{\partial}{\partial r} \left(\rho r^4 \Omega U_2 \right) + \dot{\mathcal{J}}\;,
\end{equation}
where $\nu_{\mathrm{add}}$ is the additional constant viscosity. \citet{2017A&A...599A..18E} constraint this viscosity using asteroseismic measurements for the particular low-mass red giant star, KIC 7341231, and obtained the range $\nu_{\mathrm{add}} = 10^3 - 1.3\times10^4\;\mathrm{cm}^2 \mathrm{s}^{-1}$.
This range was estimated for a model including advective transport by meridional currents and diffusive transport by shear instabilities. We use this range as a guide for the viscosity of this unknown mechanism in our models which also include transport of angular momentum by mixed modes.

\subsubsection{Evolution of rotation rates with $\nu_{\mathrm{add}}$}

\par In Figure~\ref{fig:omega_evolution_diffnu} we show the evolution of the rotation rate when taking into account the additional viscosity term. It is evident that this additional viscosity, whose value has been constrained by observations, brings models without the mixed mode mechanism closer to observations. In particular, the models with the highest additional viscosity (i.e., blue and orange lines in Fig.~\ref{fig:omega_evolution_diffnu}) are now overlapping with observations early on the RGB. Whereas for the model with the lowest additional viscosity (i.e., red line in Fig.~\ref{fig:omega_evolution_diffnu}), the core rotation rate still does not seem to match the observations on the early RGB.

\par The models which include the mixed mode mechanism with radiative damping are only slightly affected by this viscosity term -- the final rotation rates are slightly lower than the models without mixed modes (dotted lines in Fig.~\ref{fig:omega_evolution_diffnu}). On the other hand, the models without radiative damping are significantly affected by this viscosity term -- the final rotation rates are significantly lower than the models without mixed modes, even reaching values lower than observations for the higher viscosity (dashed lines in Fig.~\ref{fig:omega_evolution_diffnu}).

\par We show that the range defined by the models including the mixed mode mechanism with and without radiative damping increases significantly with increasing additional viscosity (see dotted and dashed line in Fig.~\ref{fig:omega_evolution_diffnu}). This wider range is now able to explain the entire spread of observed rotation rates for higher viscosities. The wider range also highlights the need to constrain the mixed mode amplitudes and damping rates. These constraints would allow us to narrow down the range of rotation rates obtained when including the mixed mode mechanism in our calculations, and precisely quantify the spin down of red giant cores due to mixed modes. The magnitude of the viscosity seems to have no impact on the onset of the core spin down (i.e., it remains around $7\;\mathrm{R}_{\odot}$ for all models in Fig.~\ref{fig:omega_evolution_diffnu}). This points towards the need for this complementary/additional mechanism to become efficient earlier in the RGB, with $\nu_{\mathrm{add}}$ at least higher than $5\times10^3\;\mathrm{cm}^2\mathrm{/s}$ to overlap with the observations (e.g., mechanisms including magnetic fields).

\subsubsection{Effect on the rotation profiles with $\nu_{\mathrm{add}}$}

\par Finally, we discuss the effect on the rotation profile when including an additional viscosity term. The bumps that we have seen in the rotation profiles in Section~\ref{sec:Effect in the rotation profiles} due to the contracting core are now smoothed out in all our models in Figure~\ref{fig:rotprofile_diffnu}. This is the case for all the viscosity values tested and for both models with and without mixed modes. This was expected since the role of this viscosity term is to redistribute the angular momentum between adjacent shells -- resulting in a smoothing of the steep gradients in the rotation profile. 

\par The models with the mixed mode mechanism and damped amplitudes no longer show steep gradients in the rotation profile (dotted lines in Fig.~\ref{fig:rotprofile_diffnu}) in comparison with the models without the additional viscosity (dotted lines in Fig.~\ref{fig:rotprofile}) -- indicating that the values we chose for the viscosity term were enough to efficiently redistribute the localised spin down due to the mixed modes. In contrast, in the rotation profiles of the models without radiative damping we see that despite the smoothing effect of the additional viscosity, they still have some steep gradients around the hydrogen-burning shell and at the chemical discontinuity from the first dredge-up (dashed lines in Fig.~\ref{fig:rotprofile_diffnu}). Even higher values for the viscosity term or/and an additional advection term in Equation~\ref{eq:AM__} would be required to homogenize these rotation profiles, which might not be realistic.

\par In conclusion, the inclusion of an additional viscosity term does indeed redistribute the localised spin down due to the mixed mode mechanism, therefore increasing its efficiency. Also, with this additional viscosity term the helium core is now rotating uniformly which is in better agreement with the results in literature \cite{2017MNRAS.464L..16K,2018ApJ...862....9D}. This strengthens the argument that the mixed modes along with an additional mechanism of angular momentum transport could explain the observed rotation rates.

\section{Conclusions}
\label{sec:6}

\par We have implemented angular momentum transport by mixed modes for evolved low-mass stars as derived by \citet{2015A&A...579A..31B,2015A&A...579A..30B}, which remained untested in stellar evolution codes, until now. For this purpose, we developed a post-processing code that solves the angular momentum transport equation including advection by meridional currents, diffusion by shear-induced turbulence and a source term including the mixed modes. With stellar structure models previously computed with MESA and oscillation modes from GYRE as an input in our code, we computed the rotation rates and rotation profiles of models starting from the subgiant phase until the luminosity bump on the RGB. Our results show that the spin down of red giants cores due to the mixed mode mechanism takes place later in the RGB (i.e., as of $\mathcal{N}=8$). This confirms that another mechanism is slowing the core at the end of the subgiant phase/early red giant phase.

\par Later in the RGB, the effect of the mixed mode mechanism on the rotation profile is strongly localised around the hydrogen-burning shell, the chemical discontinuity left by the first dredge-up, and in the innermost layers of the core. This localised spin down is not efficiently redistributed by the meridional circulation. Further observational information on the rotation profile may assist in probing the physical processes that dominate the angular momentum transport in red giants. We found that an additional mechanism of angular momentum transport with constant viscosity within the range $10^3 - 10^4\;\mathrm{cm}^2 \mathrm{s}^{-1}$ can diffusive the localised spin down due to mixed modes and consequently increase their efficiency in transporting angular momentum. 
Further investigation into the physical process behind this mechanism (for example magnetic instabilities) and its interaction with the mixed modes mode mechanism is beyond the scope of the current paper and will be explored in future works. 

\par The spin down of red giant cores strongly depends on the mixed modes amplitudes. Models incorporating the mixed mode mechanism, with damped amplitudes due to radiative losses, exhibit a delayed spin down with a significantly smaller magnitude than models which neglect those losses. Consequently, to accurately quantify the effect of mixed modes on the angular momentum transport it is crucial to consider the radiative damping in the mixed mode amplitudes. Current limitations include the difficulty of obtaining precise estimates of non-radial mode surface velocities from observations, as theoretical models for radiative damping rates remain unconstrained \citep[see][]{2009A&A...506...57D,2014A&A...572A..11G}.

\par Overall, the mixed modes do have a significant effect in slowing down the core of red giants. Additionally, since they have been observed in low-mass evolved stars \citep{2011Natur.471..608B,2011Sci...332..205B}, in contrast to other potential mechanisms of angular momentum transport, they should not be neglected and should be included in state-of-the-art angular momentum transport codes. Indeed, understanding the effect of this mechanism in other stages of evolution (e.g., helium core burning) is an important step. At present, however, the implementation of the mixed mode mechanism requires the computation of the mixed modes frequencies, inertias and eigenfunctions of several angular degrees and azimuthal orders for each model, which is a computationally expensive task that significantly increases the computation time of each stellar evolution track. In the future, asymptotic theory for the estimation of the mixed modes parameters \citep{2001MNRAS.328..601D} and further parameterizations for the calculation of the mixed modes momentum flux need to be explored.

\par In conclusion, the mixed mode mechanism alone cannot explain the full range of rotation rates observed in red giants. This implies that multiple mechanisms must be considered to fully explain the evolution of rotation rates along the RGB: IGWs \citep{2014ApJ...796...17F,pincon_implications_2017} or magnetic fields \citep{2015ApJ...808...35K,2019MNRAS.485.3661F,2021A&A...646A..19T,2022A&A...664L..16E,2023A&A...673A.110M} are promising candidates. Testing potential mechanisms of angular momentum transport will allow us to constrain the currently missing physical processes, and to improve the calibration of the free parameters of the known potential mechanisms. Ultimately, this will lead to a more complete description of the angular momentum transport in the radiative regions of low-mass stars.

\begin{acknowledgements}
We thank the anonymous referee for their insightful comments and remarks that improved the manuscript. 
This project was funded by the ERC Consolidator Grant DipolarSound (grant agreement \#101000296). We thank the Klaus Tschira foundation for their support.
\end{acknowledgements}

\bibliographystyle{aa}
\bibliography{refs.bib}

\begin{appendix}

\section{Mode amplitudes including radiative damping}
\label{ap:3}

\par The angular momentum extracted by mixed modes is mostly mediated by the g-dominated modes with angular degrees higher than one \citep[see discussion in][]{2015A&A...579A..31B}. This is partly due to the mixed modes flux scaling with the azimuthal order (see Eq.~\ref{eq:j_dot}). The g-dominated modes with high angular degrees are more trapped in the core and their eigenfunctions have higher amplitudes than the p-dominated modes in the core region. For that reason, these modes are more affected by the radiative damping in the core. It is therefore crucial to take this factor into account when computing the mixed modes momentum flux.

\par Following up on Section~\ref{sec:nonradialmodeamplitudes} where we describe how we estimate the mode damping and the mode surface velocities for non-radial modes in red giants, in this section, we showcase and compare our estimates with previous works in literature \citep[e.g.,][]{2009A&A...506...57D,2014A&A...572A..11G} and with observations of mixed modes in red giant stars \citep[e.g.,][]{2012A&A...548A..10M,2017A&A...598A..62M}.

\subsection{Comparing our mode amplitudes with previous works}

\par
\begin{figure*}
    \raggedright
    \includegraphics[width=1.0\textwidth]{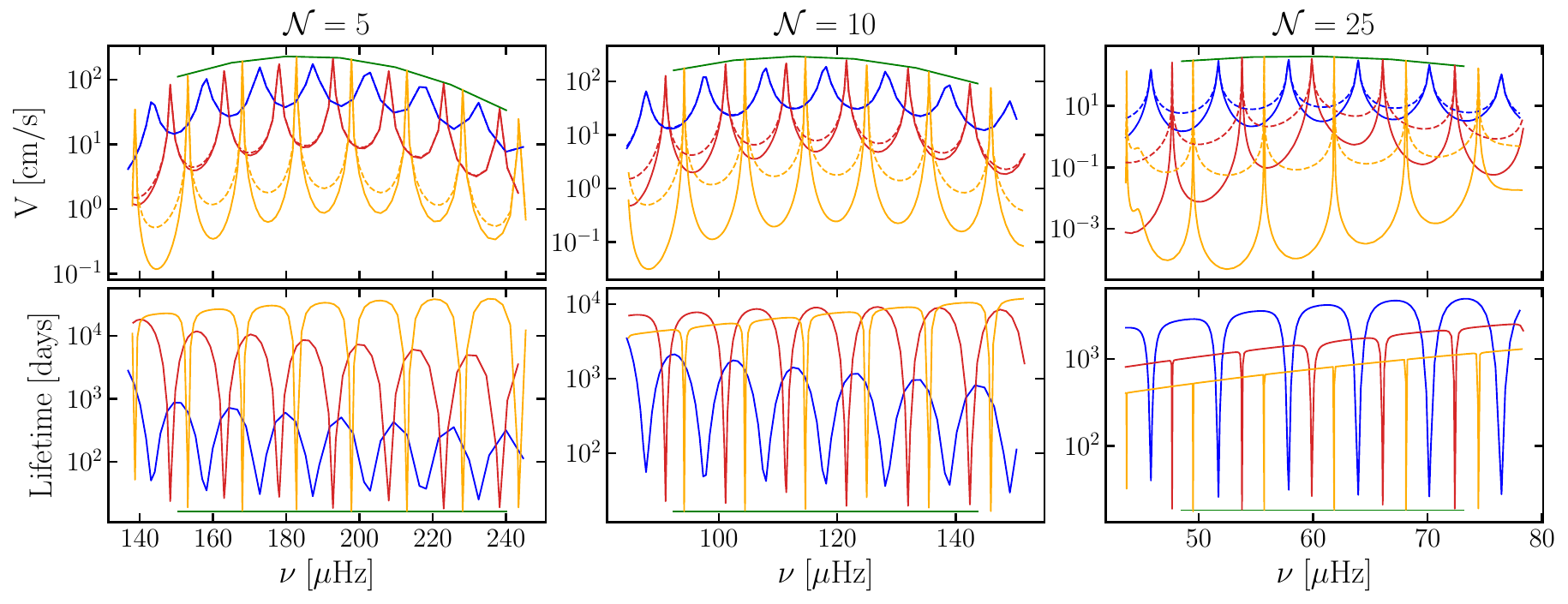}
    \vspace{-5mm}
    \caption{Top panel: Mode surface velocity (logarithmic scale) as a function of frequency for the models along a 1.0 M$_{\odot}$ track. The line-styles represent models with mode damping (solid lines) and without mode damping (dashed lines). The line colours represent the angular degrees $l=0,1,2,3$ (green, blue, red, yellow). The left, central and right panels showcase models in different evolutionary stages (with different mixed modes density). Bottom panel: Mode lifetime (logarithmic scale) as a function of frequency.}
    \label{fig:Vsurf}
\end{figure*}

\begin{figure*}
    \raggedright
    \includegraphics[width=1.0\textwidth]{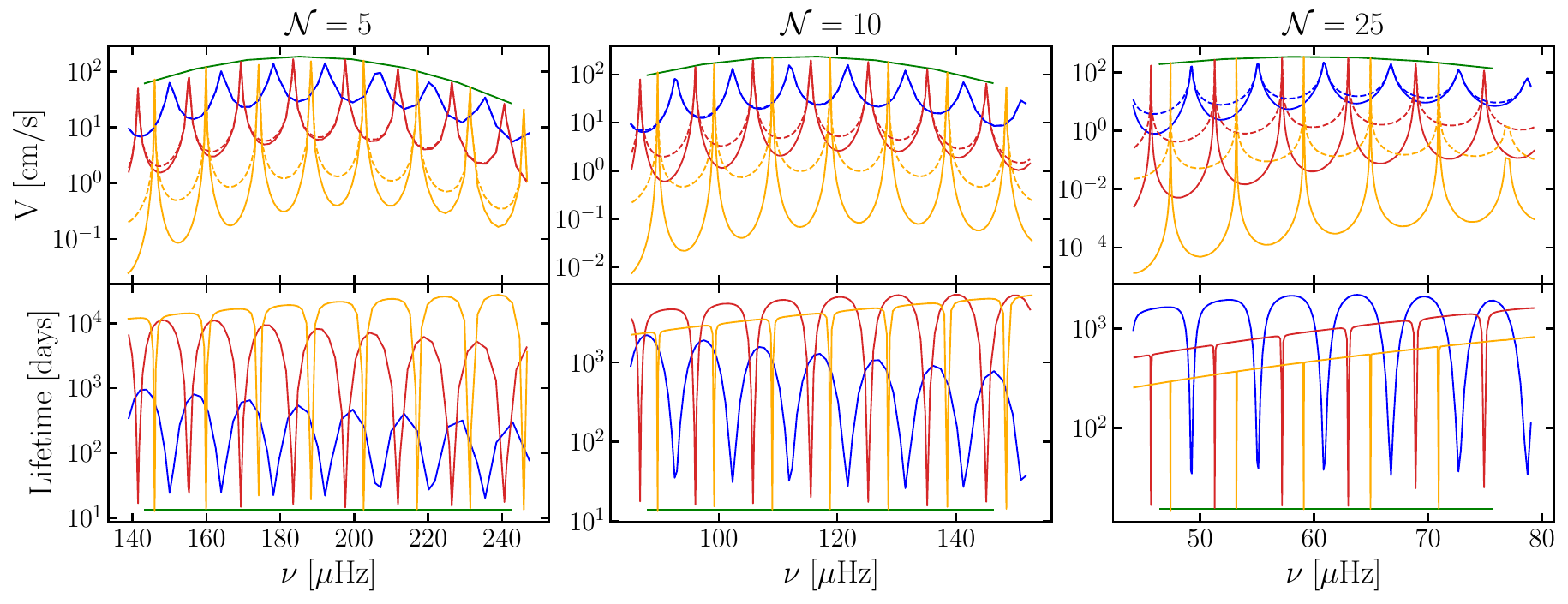}
    \vspace{-5mm}
    \caption{Same as Fig.~\ref{fig:Vsurf} for 1.3 M$_{\odot}$.}
    \label{fig:Vsurf2}
\end{figure*}

\begin{figure*}
    \raggedright
    \includegraphics[width=1.0\textwidth]{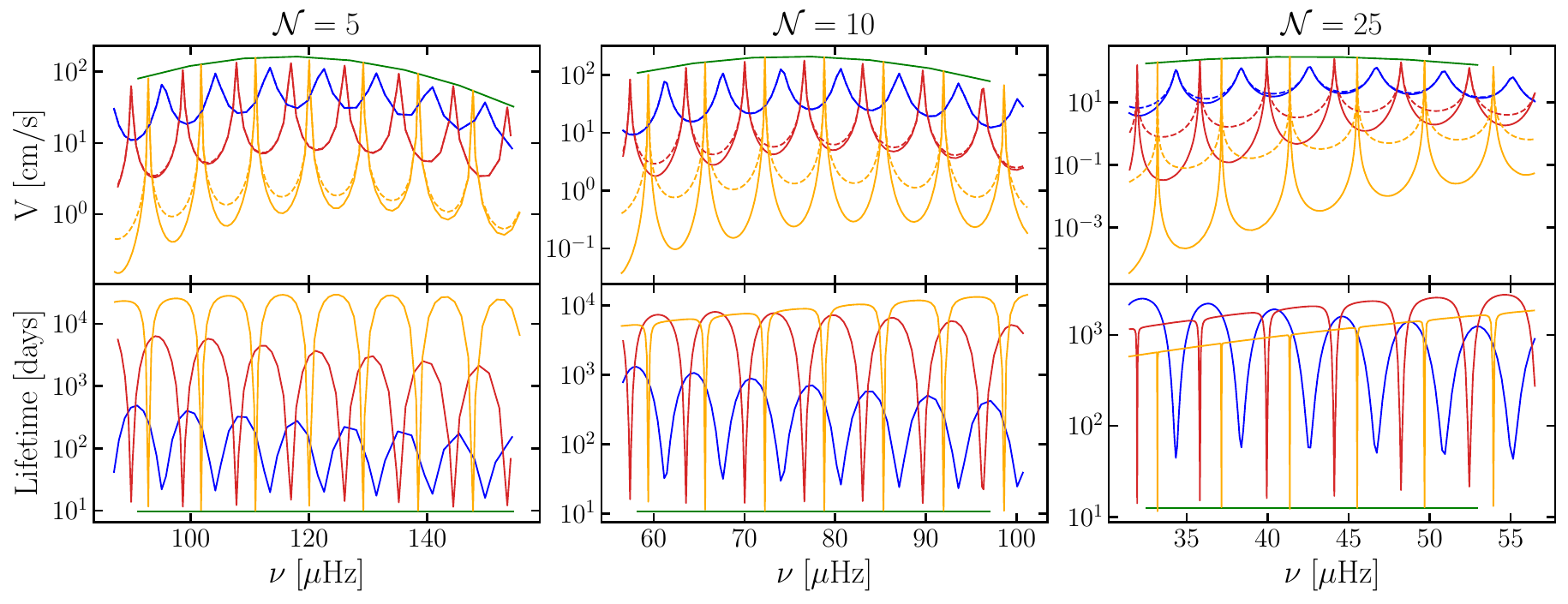}
    \vspace{-5mm}
    \caption{Same as Fig.~\ref{fig:Vsurf} for 2.0 M$_{\odot}$.}
    \label{fig:Vsurf3}
\end{figure*}

 In the top panel of Figure~\ref{fig:Vsurf}, \ref{fig:Vsurf2} and \ref{fig:Vsurf3} we show the mode surface velocity computed with and without radiative damping. As expected the radiative damping significantly impacts the g-dominated modes with higher angular degrees. For the $l=3$ modes the difference when including mode damping is around three orders of magnitude in the mode surface velocity. Since these modes mediate the transport of angular momentum, we expect a significant impact on the efficiency of the angular momentum transport mechanism (see discussion in Sect.~\ref{sec:timescales} and in App.~\ref{ap:Mixed modes momentum flux}).
It is important to note that in this work we have only computed mixed modes amplitudes for angular degrees $l=0,1,2,3$ for each model along evolution which is already computationally expensive. In future works it is however important to investigate the effect that higher angular degrees would have in the overall mixed mode momentum flux in the earlier stages of the RGB and whether this would affect the point in evolution where the mixed mode mechanism becomes efficient. For models higher on the RGB the contribution for the momentum flux of these modes with higher angular degrees would be negligible due to the higher radiative damping.
        
\par The top panels in Figure~\ref{fig:Vsurf}, \ref{fig:Vsurf2} and \ref{fig:Vsurf3} show that our estimates for the mode surface velocity of radial mode and p-dominated modes are in agreement with the surface velocities reported in \cite{2009A&A...506...57D} and in \cite{2015A&A...579A..31B}. However, the very small amplitudes of the g-dominated modes in our models cannot be constrained by theoretical models and by observations \citep[see][for details regarding the limits of mixed mode detectability]{2014A&A...572A..11G}.
        
\par The bottom panels in Figure~\ref{fig:Vsurf}, \ref{fig:Vsurf2} and \ref{fig:Vsurf3} illustrate the mode lifetimes ($\tau = 1/\eta$). We neglect the dependence of the mode lifetimes in frequency, since we obtain the mode lifetime from Equation~\eqref{eq:belkacem+12}. The order of magnitude of our radial mode lifetimes is in agreement with results from \cite{2009A&A...506...57D}. However, the mixed mode lifetimes, for the model with 1~M$_{\odot}$, are one order of magnitude smaller than predicted in \cite{2017A&ARv..25....1H}. Our mode lifetimes are also underestimated when compared with the ones reported by \cite{2009A&A...506...57D} for a model with 2~M$_{\odot}$ -- around two orders of magnitude for models at the base of the RGB, and less than one order of magnitude for models just below the bump on the RGB. The underestimation of the mode lifetimes reflects the overestimation of the radiative damping in our models which consequently decreases the mode amplitudes. This difference possibly arises from differences in the stellar models and from the different methods adopted in previous works to estimate the mode damping.

\par Summarizing, the radiative damping included in our models is slightly higher than the one predicted in previous works. Hence, we argue that our models including radiative damping should be considered as a limiting case -- lowest amplitudes predicted by scaling relations and asymptotic theory. 

\subsection{Comparing our mode amplitudes with observations}

\begin{figure*}
    \raggedright
    \includegraphics[width=1.0\textwidth]{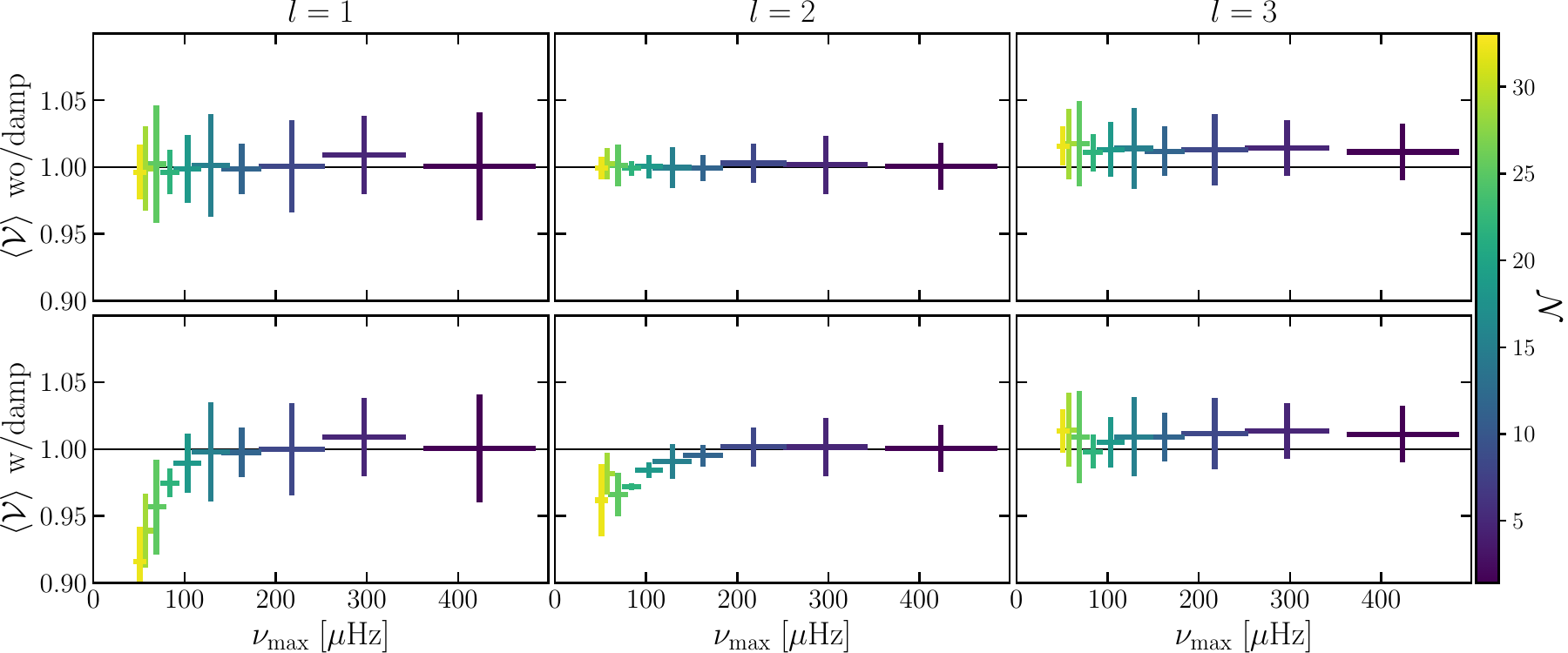}
    \caption{Top panel: Integrated mode visibilities $\left<\mathcal{V}\right>$ for models without radiative damping as a function of $\nu_{\mathrm{max}}$for the model with $1.3\;\mathrm{M}_{\odot}$. Bottom panel: Same a top panel now for models including radiative damping. The left, centre and right panels correspond to different angular degrees $l=1,2,3$, respectively. The colour of the points represents the mixed modes density of the respective models. The error bars show the standard deviation of each model (amongst all the radial orders).}
    \label{fig:visibilities}
\end{figure*}
\par In this section we investigate whether our approaches to estimate mixed modes surface velocities -- with and without radiative damping -- are in agreement with observations. 

\par For this purpose we sum the contribution of all the mixed modes at a given radial order (using Eq.~\ref{eq:V_l}) and introduce $\mathcal{V}_{l}$ as the intrinsic energy in modes of degree $l$ as a function of the energy in the radial mode,
\begin{equation}
    \mathcal{V}_{l}= \sum_{i=1}^{\mathcal{N}} \left( \frac{V_{l,i}}{V_{0,i}} \right)^2 \approx
\begin{cases}
    \sum_{i=1}^{\mathcal{N}} \left( \frac{\mathcal{M}_{0,i}}{\mathcal{M}_{l,i}} \right)^2 \frac{\eta_{0,i}}{\eta_{l,i}} \;, & \text{w/\;damp}\\
    \sum_{i=1}^{\mathcal{N}} \left(\frac{\mathcal{M}_{0,i}}{\mathcal{M}_{l,i}} \right), & \text{wo/\;damp}
\end{cases}
\label{eq:visibilities}
\end{equation}
\citep{,2014ApJ...781L..29B,2017A&A...598A..62M} where $\mathcal{N}$ is the total number of mixed modes per radial order previously defined in Equation~\eqref{eq:mixedmodesdensity}. $V_{l,i}$ corresponds to the surface velocity of the \textit{i}-th mixed mode with angular degree $l$ and $V_{0,i}$ corresponds to the surface velocity of the radial mode interpolated to the frequency of the \textit{i}-th mixed mode.  

\par As derived by \cite{2017A&A...598A..62M}, energy equipartition is preserved for the mixed modes where the damping in the core is neglected -- therefore from the theoretical framework we expect $\mathcal{V}_{l} = 1$. 
This implies that when radiative damping is taken into account in the core we depart from energy equipartition and expect $\mathcal{V}_{l}\leq 1$.

\par In Figure~\ref{fig:visibilities} we show the average of the values obtained using Equation~\ref{eq:visibilities} over all the radial orders -- $\left< \mathcal{V}\right>$. As expected, the models where we neglect radiative losses have values for $\left< \mathcal{V}\right>$ around one. The dispersion caused by the uncertainties in $\left< \mathcal{V}\right>$ is due to the variation of the number of mixed modes within each radial order. On the other hand, the models where we accounted for radiative damping depart from one for lower frequencies (where the radiative damping becomes significant). This departure from one is more pronounced for the $l=1$ modes due to their stronger mixed character (stronger coupling between p- and g-modes). Whereas in the $l=3$ modes, the g-dominated modes have a very small contribution to the $\left< \mathcal{V}\right>$ with and without radiative damping (due to their very small amplitudes at the surface) when compared to the p-dominated modes. Hence, the variation in $\left< \mathcal{V}\right>$ when accounting for radiative losses in the core is very small.

\par From observations, \cite{2017A&A...598A..62M} has shown that $\left<\mathcal{V}_{\mathrm{obs}}\right> \in [0.8-1.2]$. Both our models, with and without radiative damping, are well within this range. This suggests that the approach we use in Section~\ref{sec:nonradialmodeamplitudes} to estimate the mode surface velocities by neglecting the radiative damping is not overestimating the mixed modes surface velocities beyond the observational constraints. The same argument applies to the models including radiative damping, our approach does not underestimate the mixed mode surface velocities to the point where they are below the observational limit.

\par In conclusion, this suggests that both approaches described in Section~\ref{sec:nonradialmodeamplitudes} are valid approaches to estimate the mixed modes amplitudes from an observational point of view. It is important to note that these observational constraints are only imposed on the ratios of the mode surface velocities. There is still the possibility that we underestimate/overestimate the radial mode surface velocities and consequently our non-radial mode surface velocities are overestimated/underestimated while the ratio stays within the limits. Therefore, we argue that the best approach to estimate mixed modes surface velocities lies somewhere in between these two approaches. Future efforts are required to constrain the absolute mixed modes surface velocities.

\vspace{-5mm}\section{Standard input physics}
\label{ap:1}

\par The physical inputs in our models include the following default MESA values: equation of state with a blend of the OPAL \citep{Rogers2002}, SCVH \citep{Saumon1995}, FreeEOS \citep{Irwin2004}, HELM \citep{Timmes2000}, PC \citep{Potekhin2010}, and Skye \citep{Jermyn2021}. Radiative opacities are mainly from OPAL \citep{Iglesias1993,Iglesias1996}, with low-temperature data from \citet{Ferguson2005} and the high-temperature by \citet{Poutanen2017}. Nuclear reaction rates are from JINA REACLIB \citep{Cyburt2010}, NACRE \citep{Angulo1999} and additional tabulated weak reaction rates by \citet{Fuller1985, Oda1994, Langanke2000}.

\par The initial chemical composition follows GS98 solar composition \citep{Grevesse1998}, with a metallicity of $Z=0.0188$ and a helium mass fraction of $Y = 0.275$. As an outer boundary condition we used the Eddington gray approximation. Convection is treated according to the standard mixing-length theory as presented by \cite{Cox1968PrinciplesOS} with a mixing length parameter $\alpha_{\mathrm{mlt}}=1.8$. Overshooting is implemented via a convective diffusion coefficient exponentially decaying beyond the boundary of convection regions: whether above convective cores or below convective envelopes with $f_{\mathrm{ov}}=0.015$ and $f_{\mathrm{ov},0}=0.005$. We computed the evolutionary tracks without rotation with a high temporal and spatial resolution.

\section{Mixed mode momentum flux}
\label{ap:Mixed modes momentum flux}

\par As the star evolves along the RGB, its internal structure changes and that affects the behaviour of waves propagating throughout its interior. These changes modify the regions where pressure and gravity modes propagate, they also modify their frequencies and amplitudes. All these properties play a role when estimating the efficiency of the mixed modes for angular momentum transport. 

\par In this section we discuss how the mixed modes momentum flux varies along evolution when accounting for radiative damping in the core (Sect.~\ref{ap:Momentum flux by damped mixed modes}) and the approximation we introduced to improve numerical convergence in our models (Sect.~\ref{ap:Approximation in the momentum flux profile}).

\subsection{Momentum flux by damped mixed modes}
\label{ap:Momentum flux by damped mixed modes}

\par As discussed in \cite{2015A&A...579A..31B} the increase in the mixed modes momentum flux along evolution is due to an increase in the following quantities: mixed mode density, mode amplitude, radial wave number and buoyancy frequency (see Eq.~\ref{eq:j_dot}). However, the radiative losses in the core also increase as the star evolves. To better estimate the efficiency of the mixed mode mechanism we compare models with and without the effect of radiative damping on the mixed mode amplitudes. 

\par
\begin{figure}
   \raggedright
   \includegraphics[width=0.5\textwidth]{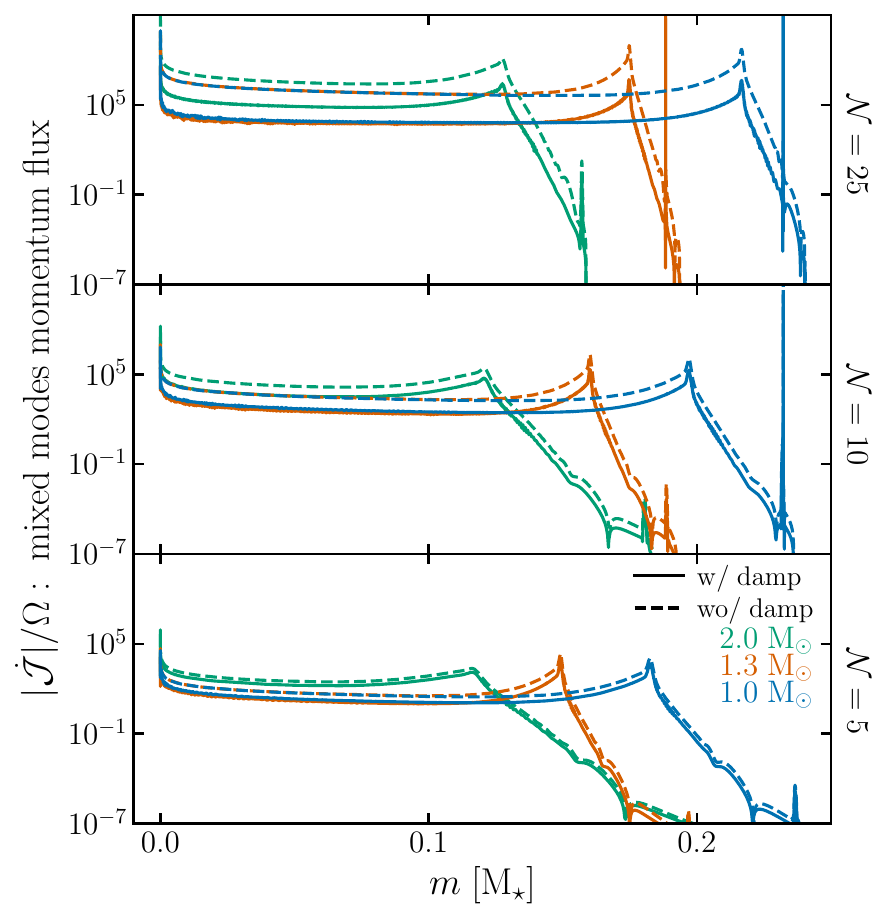}
   \caption{Absolute value for the mixed modes momentum flux normalized by the rotation rate (logarithmic scale) as a function of fractional mass. The different colours represent the mass of the models. The different panels show models at different evolutionary stages: just below the RGB luminosity bump (top panel), lower on the RGB (central panel), and at the base of the RGB (bottom panel). The mixed modes momentum flux was computed with mode damping (filled line) and without mode damping (dashed line).}
    \label{fig:jdot_allmasses}
\end{figure}

 For the models early on the RGB, we find a small difference in the mixed modes momentum flux when including radiative damping (see bottom and middle panel in Fig.~\ref{fig:jdot_allmasses}). Just below the RGB bump, the models with mode damping show a mixed modes momentum flux around two orders of magnitude smaller than the models without mode damping (see upper panel in Fig.~\ref{fig:jdot_allmasses}). This translates into a significant decrease in the efficiency of the mixed mode mechanism. Overall, the amount of radiative damping affecting $\dot{\mathcal{J}}$ seems to be independent of stellar mass.

\par The mixed modes momentum flux $\dot{\mathcal{J}}$ is higher in the very centre of the star, in the region around the hydrogen-burning shell and around spikes in the chemical discontinuity region due to the first dredge-up. This is the case for all models along the RGB independent of stellar mass (see Fig.~\ref{fig:jdot_allmasses}). In the very centre, the momentum flux is higher due to the increase in the amplitude of the radial eigenfunction. The density also plays a role, however, when solving the angular momentum transport equation we divide by the density so that dependency disappears. In the other regions, the momentum flux is obviously increased due to the buoyancy frequency dependency. Given that the mixed modes momentum flux is computed by summing all modes within a frequency range, the spatial shift between the radial eigenfunctions of consecutive modes causes a smoothing effect in the profile of $\dot{\mathcal{J}}$ \citep[also seen by][]{2015A&A...579A..30B}. We find that the smoothing effect is reduced for models with mode damping (compare orange lines in both panels in Fig.~\ref{fig:timescales_}).

\par The mixed modes frequencies, amplitudes and propagation regions vary for different stellar masses, and therefore $\dot{\mathcal{J}}$ is expected to be different for different stellar evolutionary tracks.
Figure~\ref{fig:jdot_allmasses} shows negligible differences between the 1 M$_{\odot}$ and 1.3 M$_{\odot}$ models. The most important difference is the peak of $\dot{\mathcal{J}}$ around the hydrogen-burning shell which appears to be slightly smaller for the 1 M$_{\odot}$ model.
On the other hand, there is a significant difference when comparing with the $2.0\;\mathrm{M}_{\odot}$ model. The $\dot{\mathcal{J}}$ (with and without damping) reaches higher values in the shells below the hydrogen-burning shell (around two orders of magnitude) and slightly smaller values than the 1.3 M$_{\odot}$ model at the hydrogen-burning shell. This translates into a less localised spin down of the core for the $2\;\mathrm{M}_{\odot}$ model compared with the other masses due to the higher density and due to the buoyancy frequency profile.

\begin{figure}
   \raggedright
   \includegraphics[width=0.5\textwidth]{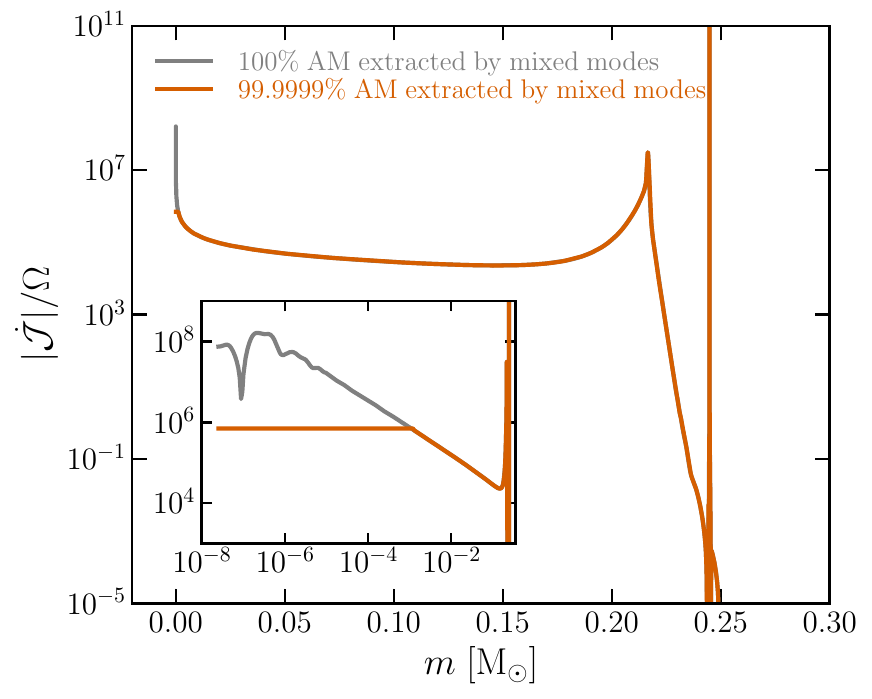}
   \caption{Absolute value for the mixed modes momentum flux normalized by the rotation rate (logarithmic scale) as a function of fractional radius for a $1.3\;\mathrm{M}_{\odot}$ model high in the RGB. The original mixed modes flux computed using Eq.~\ref{eq:j_dot} is represented by the grey line. The modified mixed modes flux is represented by the orange line. The smaller panel shows a close up of the very central region of the star where the flux was modified.}
    \label{fig:jdot_approx}
\end{figure}

\subsection{Approximation in the momentum flux profile}
\label{ap:Approximation in the momentum flux profile}

\par The momentum flux represented in Figure~\ref{fig:jdot_approx} is overall negative, meaning that angular momentum is being extracted from the rotating shells of the star. However, closer to the inner region of the star there is a sign flip in the mixed modes momentum flux, where it switches from negative values to positive values (interpreted as momentum being deposited in the shell). This sign flip happens below fractional mass $10^{-3}$~M$_{\odot}$ due to the sign flip in the radial wave number which corresponds to a region where the mixed modes no longer propagate, and their amplitudes decay exponentially.

\par The region in the very centre of the star, where the mixed modes flux gradient increases and sign flip, constitutes a very small fraction of mass of the star which makes the convergence of the numerical scheme very challenging. One solution would be to increase the mesh resolution around that region to ensure the numerical scheme can relax within the maximum number of iterations. Our automated adaptative mesh scheme already increases the number of mesh points in that region, however, an unfeasible amount of mesh points is needed in order to fully resolve it, which increases significantly the computation time. Figure~\ref{fig:jdot_approx} shows the mixed modes momentum flux of a chosen model with the purpose of illustrating the modification performed to ensure convergence of the numerical scheme. The modification implemented reduces the $\dot{\mathcal{J}}$ gradient and erases the sign flip, with a negligible loss of angular momentum of $0.0001\%$. Surprisingly, the gradient caused by the spike from the chemical discontinuity around fractional mass $0.24\;\mathrm{M}_{\odot}$ is well resolved by the numerical scheme since it appears only for more evolved models, and it moves inwards in the mass shells.

\clearpage
\begin{figure*}
\begin{multicols}{2}
\section{Timescales associated with angular momentum transport: 1 M$_{\odot}$ and 2 M$_{\odot}$}
\label{ap:Timescales}
\par In this section we adopt the same approach as in Section~\ref{sec:timescales} and briefly discuss the timescales of the 1 M$_{\odot}$ and 2 M$_{\odot}$ models for completeness.
\subsection{Timescales within the radiative region}
\par In Figure~\ref{fig:timescales1M} and \ref{fig:timescales2M} we show the timescales associated with the angular momentum transport mechanisms at $\mathcal{N}=30$.
\par Similarly to the 1.3 M$_{\odot}$, the mixed modes are more efficient in the innermost shells, in the hydrogen-burning shell and in the chemical discontinuity due to the first dredge-up. As expected from the discussion in the Appendix~\ref{ap:Momentum flux by damped mixed modes}, in the 2 M$_{\odot}$ model the mixed modes are less efficient in the hydrogen-burning shell than for the other masses (see green line in Fig.~\ref{fig:timescales2M}) which results in a less localised spin down. This can be seen by comparing the mixed mode timescale (green line in Fig.~\ref{fig:timescales2M}) at the hydrogen-burning shell (labelled as H-shell) and at the innermost point in the star. Whilst for the 1 M$_{\odot}$ and 1.3 M$_{\odot}$ models, in these two regions $\tau_{\mathrm{modes}}$ is roughly the same order of magnitude, for the 2 M$_{\odot}$
$\tau_{\mathrm{modes}}$ is around two orders of magnitude higher in the hydrogen-burning shell than in the innermost point.
\subsection{Timescales evolution along the RGB}
\par Comparing the timescale value at the hydrogen-burning shell for the 1 M$_{\odot}$ (see Fig.~\ref{fig:timescales_1M}) with the 1.3 M$_{\odot}$ (see Fig.~\ref{fig:timescales_}), it suggests that the points in evolution where the mixed mode mechanism becomes efficient are very similar (i.e., around $\mathcal{N}=8$ without mode damping and $\mathcal{N}=14$ with mode damping). This seems to be a combination of slightly less efficient contraction (as expect due to the lower mass) and of slightly less efficient transport of angular momentum by the mixed modes (as discussed in Ap.~\ref{ap:Momentum flux by damped mixed modes}).
On the other hand, in the 2 M$_{\odot}$ model (see Fig.~\ref{fig:timescales_2M}) the points in evolution where the mixed modes become efficient are significantly different (i.e., around $\mathcal{N}=15$ without mode damping and $\mathcal{N}>35$ with mode damping). This is due to the more efficient contraction (due to the higher mass). For this model the meridional circulation is also significantly more efficient than for the other masses (light grey line in Fig.~\ref{fig:timescales_2M}).

\par After comparing models with different stellar masses it is clear that the mass of the models affects the efficiency of the mixed mode mechanism. Nonetheless, these three models with different masses are not sufficient to make conclusions regarding the mass dependency on the mixed mode efficiency, particularly, since it depends on the ratio between the timescales associated with the angular momentum transport of all mechanisms involved.
\end{multicols}
\vspace{-6mm}
    \centering
    \begin{multicols}{2}
    \includegraphics[width=0.49\textwidth]{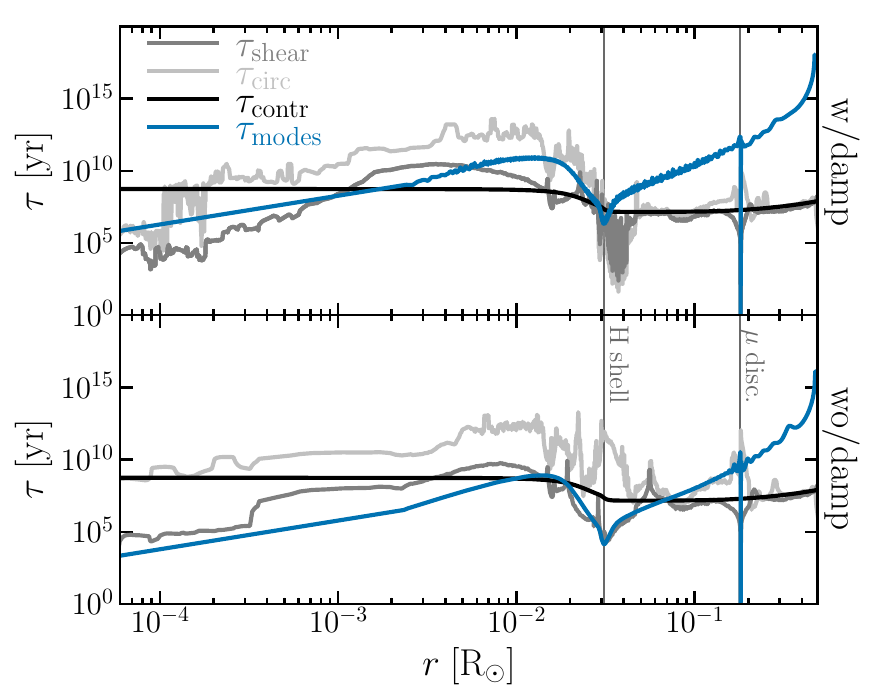}
    \vspace{-7mm}\caption{Same as Fig.~\ref{fig:timescales} for 1 M$_{\odot}$.}
    \label{fig:timescales1M}
    \includegraphics[width=0.49\textwidth]{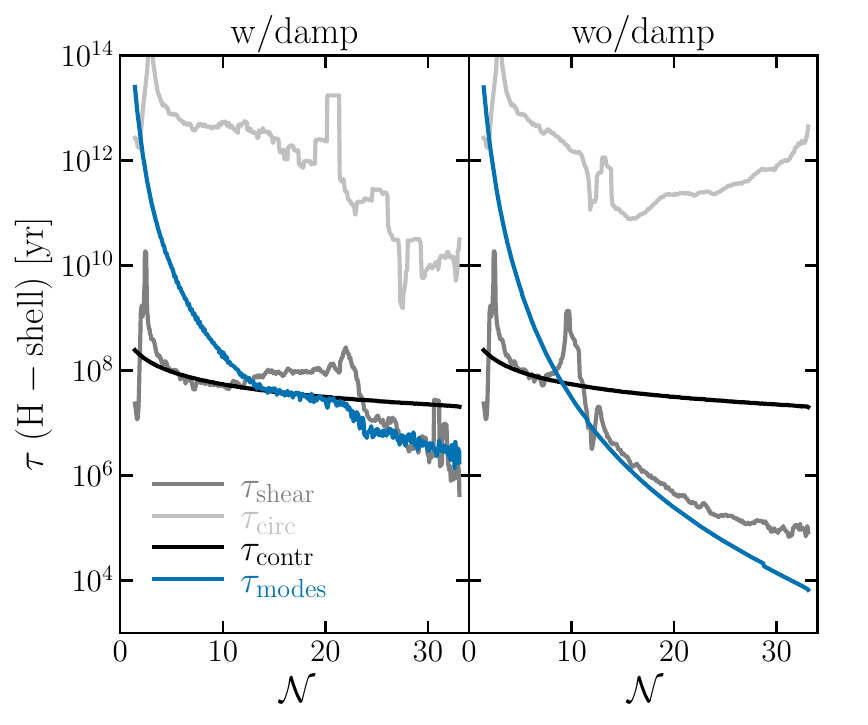}
    \vspace{-7mm}\caption{Same as Fig.~\ref{fig:timescales_} for 1 M$_{\odot}$.}
    \label{fig:timescales_1M}
    \includegraphics[width=0.49\textwidth]{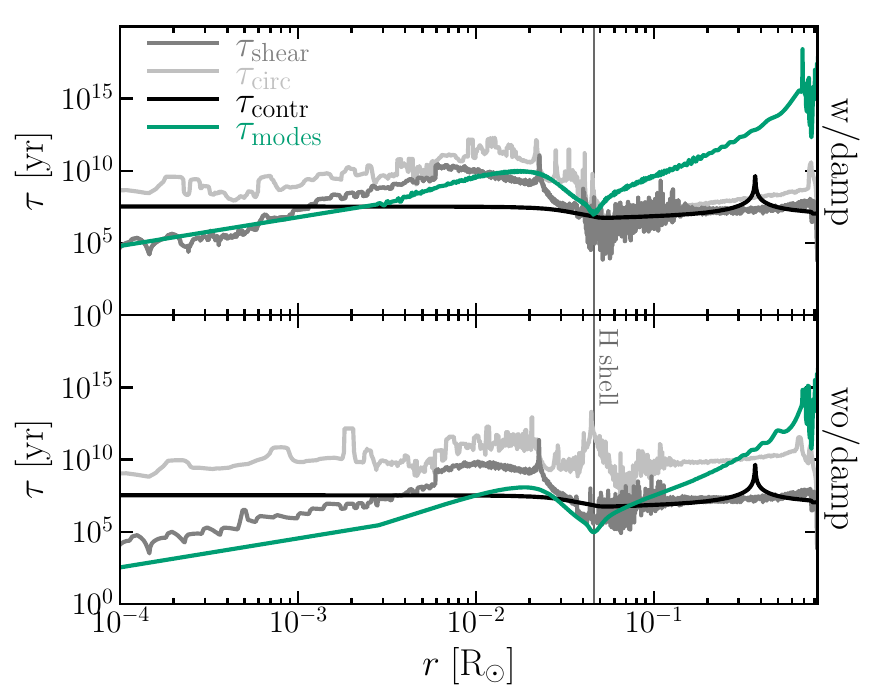}
    \vspace{-7mm}\caption{Same as Fig.~\ref{fig:timescales} for 2 M$_{\odot}$.}
    \label{fig:timescales2M}
    \includegraphics[width=0.49\textwidth]{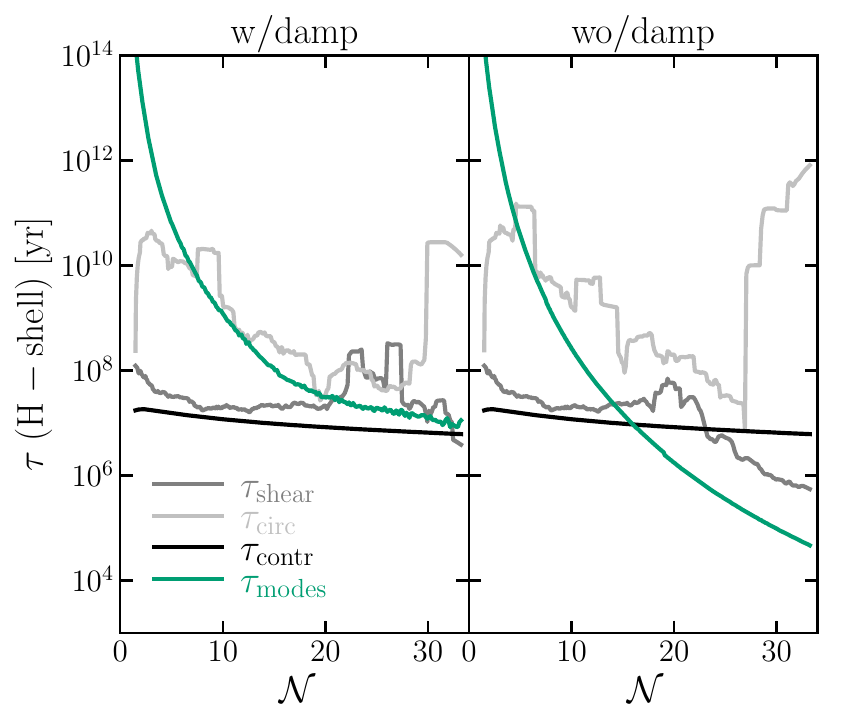}
    \vspace{-7mm}\caption{Same as Fig.~\ref{fig:timescales_} for 2 M$_{\odot}$.}
    \label{fig:timescales_2M}
    \end{multicols}
\end{figure*}

\begin{figure*}
\begin{multicols}{2}
\section{Comparing our models with MESA rotating models}
\label{ap:MESA} 
\par In this section we provide a comparison between the rotation profiles in the models computed using our post-processing code (hereafter \textit{AM\;code}) and the rotation profiles in the models from MESA (hereafter \textit{AM\;MESA}) for the case where the contracting shells dominate the angular momentum redistribution.

\par All the models in this section were obtained using the release version 23.05.1 of MESA \citep{paxton_modules_2011, paxton_modules_2013, paxton_modules_2015, paxton_modules_2018, paxton_modules_2019,Jermyn2023}. We computed stellar evolution tracks for three stellar masses 1.0, 1.3 and 2.0 M$_{\odot}$ at solar metallicity with the standard physical inputs described in Appendix~\ref{ap:1}. The implementation of rotation in our code is, however, different from the implementation in MESA. In our post-processing code we adopt an advective-diffusive approach to solve the angular momentum transport equation (see Sect.~\ref{sec:rotmodels} for full details). For comparison purposes we only include angular momentum transport by meridional currents and by shear-induced turbulence in these models. On the other hand, MESA adopts a diffusive only approach to solve the angular momentum transport equation while simultaneously solving the equations for the stellar structure. In the MESA models, rotation was included following Equation~B4 described in \cite{paxton_modules_2013}, with the turbulent viscosity coefficient ($\nu$) determined only by the kinematic shear viscosity coefficient. All the models were initialized with uniform initial rotation profile $\Omega_{\mathrm{ini}}=1.57\times10^{-6}$ rad/s at the base of the RGB.

\par We show that the rotation profiles obtained using our code reproduce the rotation profiles obtained using MESA reasonably well (see Fig.~\ref{fig:MESAcomp}). Despite the different numerical implementations and the different angular momentum transport mechanisms included, both models yield similar results because the dominant mechanism in both cases is the contraction of the shells.
This is the case for all the three masses and for models along the RGB. Some differences arise at the boundary between the radiative and convective regions that might be related to the different approaches implemented for the angular momentum transport and the missing feedback into the stellar structure in our code. Nonetheless, the features we find in our models, the bumps due to contraction of the core and the step at the bottom of the convection region (discussed in Sect.~\ref{sec:Effect in the rotation profiles}), are also present in the MESA rotating models.
\end{multicols}

    \centering
    \includegraphics[width=1\textwidth]{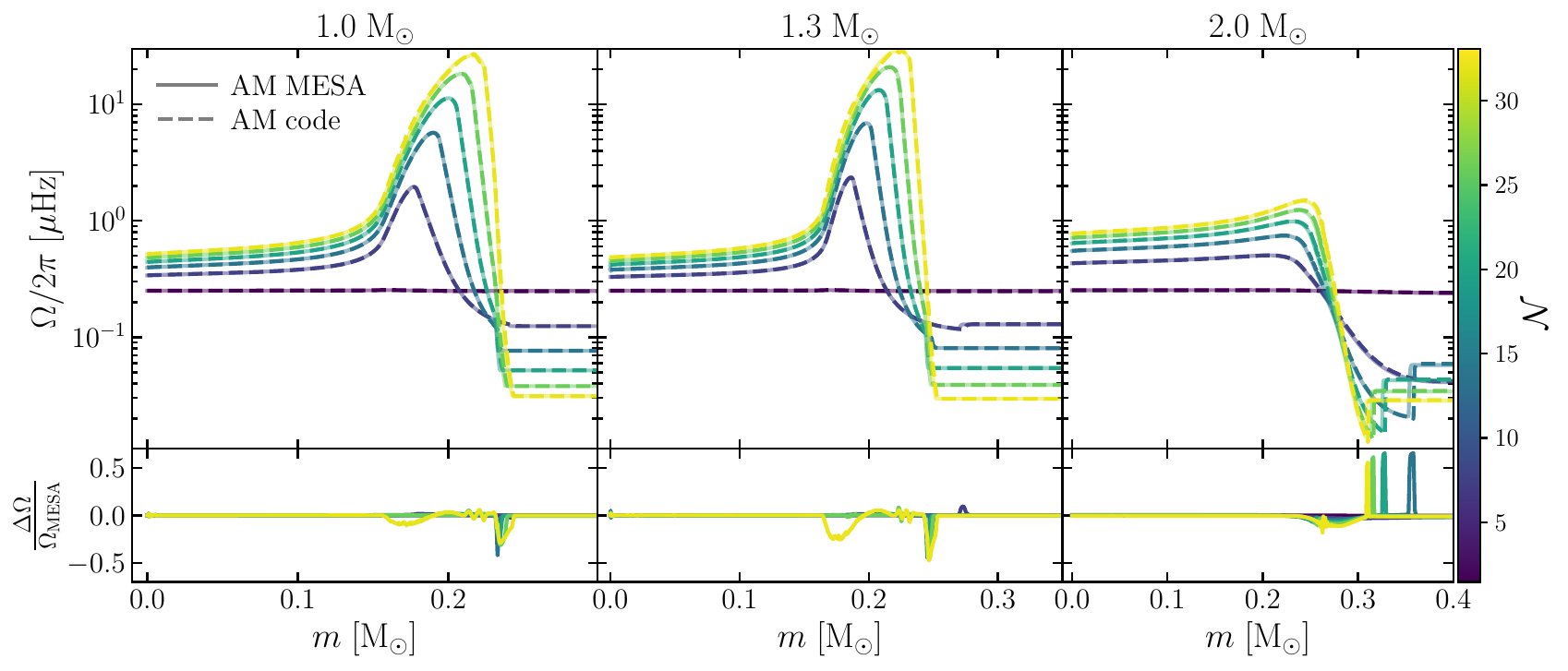}
    \caption{Top panel: Rotation rate (logarithmic scale) as a function of the fractional stellar mass. The line colour represents the evolutionary stage of the model in terms of mixed mode density. The dashed lines represent models computed using MESA angular momentum transport prescription. The solid lines represent models computed using our angular momentum transport prescription. The left, centre and right panels showcase models with different stellar masses. Bottom panel: Relative difference in rotation rate between MESA and our models.}
     \label{fig:MESAcomp}
\end{figure*}

\end{appendix}

\end{document}